\documentclass[usenatbib,fleqn,useAMS]{mnras}

\pdfoutput=1
\usepackage{hyperref,graphicx,natbib,times,amssymb,amsmath,pstricks,soul,color, epstopdf}
\usepackage{newtxtext,newtxmath,ulem}
\usepackage{aas_macros}
\usepackage{bm,url}
\usepackage{textcomp}

\usepackage{cleveref}
\usepackage[T1]{fontenc}
\usepackage{ae,aecompl}
\usepackage{mathtools}

\newcommand  *{\B}[1] {\boldsymbol{#1}}

\title[Traffic Jams in Inclined Circumbinary Discs]{Dust Traffic Jams in Inclined Circumbinary Protoplanetary Discs I. Morphology and Formation Theory}
\author[H. Aly et al.]{
Hossam Aly,$^{1,2}$\thanks{E-mail: hossam.saed@gmail.com}
Jean-François Gonzalez,$^{1}$,
Rebecca Nealon$^{3,4}$,
Cristiano Longarini$^{5}$,
\newauthor
Giuseppe Lodato$^{5}$,
and Daniel J. Price$^{6}$
\\
        $^{1}$ Univ Lyon, Univ Claude Bernard Lyon 1, Ens de Lyon, CNRS, Centre de Recherche Astrophysique de Lyon UMR5574, F-69230, Saint-Genis-Laval, France\\
        $^{2}$ Zentrum für Astronomie der Universität Heidelberg, Astronomisches Rechen-Institut, Mönchhofstr. 12-14, 69120 Heidelberg\\
        $^{3}$ Centre for Exoplanets and Habitability, University of Warwick, Coventry CV4 7AL, UK\\
        $^{4}$ Department of Physics, University of Warwick, Coventry CV4 7AL, UK\\
        $^{5}$ Universit\`a degli Studi di Milano, Via Giovanni Celoria 16, 20133 Milano, Italy\\
        $^{6}$ School of Physics and Astronomy, Monash University, Clayton, VIC 3800, Australia\\
}       
\date{Accepted  .
      Received ;
      In original form ;
      }
\pagerange{\pageref{firstpage}--\pageref{lastpage}}
\pubyear{2021}
\begin{document}
\label{firstpage}
\pagerange{\pageref{firstpage}--\pageref{lastpage}}
\maketitle
\begin{abstract}
Gas and dust in inclined orbits around binaries experience precession induced by the binary gravitational torque. The difference in precession between gas and dust alters the radial drift of weakly coupled dust and leads to density enhancements where the radial drift is minimised. We explore this phenomenon using 3D hydrodynamical simulations to investigate the prominence of these `dust traffic jams' and the evolution of the resulting dust sub-structures at different disc inclinations and binary eccentricities. We then derive evolution equations for the angular momentum of warped dust discs and implement them in a 1D code and present calculations to further explain these traffic jams. We find that dust traffic jams in inclined circumbinary discs provide significant dust density enhancements that are long lived and can have important consequences for planetesimal formation.
\end{abstract}



\section{Introduction}
The theory of planet formation through `core accretion' \citep{Safronov1969, Goldreich&Ward1973} posits that dust grows through collision and coagulation from micron scales to kilometer-size planetesimals. One of the main challenges to this picture is the `radial drift barrier' \citep{Whipple1972, Weidenschilling1977}. The pressure gradient in the disc causes the gas component to orbit at sub-Keplerian velocities, while the dust is not affected by gas pressure. This causes dust particles to experience a headwind due to the drag force exerted by the gas and drift towards the central star. This is best quantified by the Stokes number St:
\begin{equation}
    \mathrm{St}\equiv\Omega_\mathrm{k} \tau_\mathrm{s}=\frac{\upi}{2}\frac{\rho_\mathrm{s} s}{\Sigma_\mathrm{g}},
\end{equation}
where $\Omega_\mathrm{k}$ is the Keplerian angular speed, $\tau_\mathrm{s}$ is the characteristic stopping time, $\rho_\mathrm{s}$ is the dust intrinsic density, $\Sigma_\mathrm{g}$ is the gas surface density, and $s$ is the particle size. The above expression is valid when assuming Epstein drag \citep{Epstein1924}. Dust particles with St $\sim 1$ are affected by radial drift the most.

Observationally, we know that discs solve the radial drift problem and manage to trap enough dust particles to form planets. This is evident from the large number of exoplanets detected, as well as the recent detection of ongoing planet formation, for example through kinematic signatures (e.g., \citealt{PinteEtal2020}) or in the case of the PDS 70 system through direct detection \citep{ChristiaensEtal2019,IsellaEtal2019,BenistyEtal2021}. Moreover, recent advances in observation techniques, particularly through the Atacama Large Millimeter Array (ALMA) and the SPHERE instrument on the Very Large Telescope (VLT), have recently provided a wealth of observations showing sub-structures in protoplanetary discs. These sub-structures can be axisymmetric gaps or rings (e.g., \citealt{ALMA2015,HuangEtal2018a,FedeleEtal2018,LongEtal2018}) or non-axisymmetric (e.g., \citealt{MutoEtal2012,VanDerMarelEtal2013,GarufiEtal2013,BenistyEtal2015}). 

A number of theoretical models have been proposed to solve the radial drift barrier and explain the formation of sub-structures in protoplanetary discs. Most of these models rely on the idea that a `dust trap' forms at gas pressure maxima as a result of vanishing drift velocities \citep{NakagawaEtal1986}. Proposed mechanisms for creating such dust traps include vortices \citep{Barge&Sommeria1995,Klahr&Henning1997}, spirals formed as a result of gravitational instability (but not spirals induced by planets) \citep{RiceEtal2005,DipierroEtal2015}, gaps induced by embedded planets \citep{FouchetEtal2010,GonzalezEtal2012,PinillaEtal2012}, and self-induced dust traps \citep{GonzalezEtal2017,VericelEtal2020}.

Since many stars form in binary systems, the same process will lead to formation of circumbinary discs. The fact that some of these discs are able to form planets is confirmed by recent observations of circumbinary planets \citep{DoyleEtal2011,WelshEtal2012,OroszEtal2012a,OroszEtal2012b,WelshEtal2015,KostovEtal2013,KostovEtal2016, OroszEtal2019}. Interestingly, circumbinary discs have been suggested recently as explanations for observed sub-structures in so-called `transition' discs (e.g., HD 142527 \citealt{PriceEtal2018a}, IRS 48 \citealt{CalcinoEtal2019}, AB Aurigae \citealt{PobleteEtal2020}, CrA-9 \citealt{ChristiaensEtal2021}, TWA 3A \citealt{CzekalaEtal2021}, and HD 143006 \citealt{BalabioEtal2021}).

The process of dust evolution is even less understood in the context of circumbinary discs. While the relative rarity of observed circumbinary exoplanets is often attributed to observational difficulties \citep{EggenbergerEtal2007,Martin&Triaud2014}, there are theoretical reasons to suspect that planet formation might face even more challenges than in discs around single stars. Perturbations from the binary can increase planetesimals eccentricity and result in orbital crossings and higher impact velocities, inhibiting further growth \citep{PaardekoopeEtal2012,Silsbee&Rafikov2021}. Disc lifetimes might be shorter around binaries, as reported by \citet{KrausEtal2012} for the Taurus-Auriga star forming region. Indeed, \citet{Alexander2012} found that photoevaporation is likely responsible for this shortened lifetime for discs around binaries with relatively wide separations. There is, however, observational evidence that lifetimes of circumbinary discs can be extended: HD98800 B, V4046 Sgr, and AK sco \citep{SoderblomEtal1998,Mamajek&Bell2014,CzekalaEtal2015}.

Star formation from molecular cloud cores collapse is a chaotic process that is expected to result in initially misaligned and warped discs \citep{Bonnell&Bastien1992,OffnerEtal2010,Bate2010,Bate2018}. We expect this to be the general case for circumbinary discs as well as circumstellar ones. Indeed, some misaligned circumbinary discs have already been reported (e.g., KH 15D \citealt{Chiang&Murray2004,WinnEtal2004,Lodato&Facchini2013,SmallwoodEtal2019,Fang2019,PoonEtal2021}, GG Tau A \citealt{Khoeler2011,AndrewsEtal2014,AlyEtal2018}, IRS 43 \citealt{BrinchEtal2016}, L1551 NE \citealt{TakakuwaEtal2017}, and HD 98800B \citealt{Kennedy2019}). Misaligned discs around binaries experience a gravitational torque that leads to radially differential precession, which causes disc warping and twisting. For the gas component, warps are expected to propagate in a wave-like manner in thicker discs with low viscosity that are more relevant in protoplanetary contexts \citep{Papaloizou&Lin1995,Nelson&Papaloizou1999,FacchiniEtal2013}, as opposed to the diffusive warp propagation that occurs in more viscous, thinner discs expected around black holes \citep{Papaloizou&Pringle1983,Ogilivie1999,Lodato&Pringle2007,LodatoPrice2010}. The final state of gas discs precessing around binaries depends on the binary parameters; for circular and low eccentricity binaries, the disc will align with the binary plane, either in a prograde or retrograde sense depending on the initial misalignment \citep{KingEtal2005,Nixonetal2011}. For eccentric binaries, discs with high initial misalignments will align in a polar configuration around the binary eccentricity vector \citep{Farago&Laskar2010,AlyEtal2015,Martin&Lubow2017,Zanazzi&Lai2018}. Either way, the disc is susceptible to breaking when the binary torque is larger than the viscous torque and the disc communicates the precession more slowly than it occurs \citep{NixonEtal2013,DoganEtal2018}.

While gas evolution in misaligned circumbinary discs has been studied in detail, the behaviour of dust in these conditions has only recently started to be investigated. \citet{Aly&Lodato2020} performed 3D Smoothed Particle Hydrodynamics (SPH) simulations of misaligned circumbinary discs of gas and dust and found that dust with St $\geq 10$ precesses almost independently from the gas component. While gas precesses almost rigidly with constant frequency after an initial period of twist propagation, the dust tends to precess with a radially differential dependence similar to that of a test particle. This difference in precession profiles alters the structure of the dust disc and forms rings. \citet{LongariniEtal2021} investigated this phenomenon analytically and numerically and showed that geometrical projections of the dust onto the gas plane leads to a prediction of two radii where the dust piles up. If the dust component is uncoupled from the gas, the effects of gas drag on the dust precession profile can be neglected and the dust will precess with the frequency of a test particle around a binary:
\begin{equation}
    \Omega_\mathrm{p}(R)=\frac{3}{4} \frac{\sqrt{G M} \eta a^{2}}{R^{7 / 2}},
    \label{equ:precess_circular}
\end{equation}
where the binary components have masses $M_1$ and $M_2$, with $M=M_1+M_2$, $G$ the gravitational constant, $\eta = M_1M_2/(M_1+M_2)^2$, $a$ is the binary semi-major axis, and $R$ is the radius measured from the centre of mass.
The gas, however, reaches a steady state where it precesses rigidly \citep{Papaloizou&Terquem1995,LarwoodEtal1996,Larwood&Papaloizou1997,Lodato&Facchini2013} with frequency:
\begin{equation}
    \omega_\mathrm{p} = \frac{\int_{R_{\text {in }}}^{R_{\text {out }}} \Omega_\mathrm{p}(R) L_\mathrm{g}(R) 2 \upi R \mathrm{d} R}{\int_{R_{\text {in }}}^{R_{\text {out }}} L_\mathrm{g}(R) 2 \upi R \mathrm{d} R} 
\end{equation}
where $L_\mathrm{g}(R)$ is the magnitude of the specific gas angular momentum density and can be evaluated from $L_\mathrm{g}(R)=\Sigma_\mathrm{g} R^2 \Omega$.
By equating the two above expressions, \citet{LongariniEtal2021} estimated the radius at which the gas and (uncoupled) dust share the same precession frequency. This co-precession radius $R_{\text{cp}}$ can thus be computed from:
\begin{equation}
R_\text{cp} = R_\text{in} \xi^{-2/7},
\label{equ:Rcp}
\end{equation}
with the function $\xi$ being:
\begin{equation}
\xi = \frac{\int_{1}^{x_{\text {out }}} x^{-2}\sigma(x) \text{d} x}{\int_{1}^{x_{\text{out}}} x^{3 / 2} \sigma(x) \text{d} x},
\end{equation}
where $\sigma=\Sigma/\Sigma_\text{in}$ and $x=R/R_\text{in}$, with $R_\text{in}$ being the inner disc radius. At this co-precession radius, the difference between the gas and dust orbiting speeds is only due to the pressure support felt by the gas but not the dust. There exist two radii, inner and outer to the co-precession radius, where projection of the dust orbiting speed onto the gas plane will offset this reduction and the radial drift will vanish, forming dust rings. \citet{LongariniEtal2021} estimate these two radii assuming that the difference in the inclination angle between the gas and dust is minimum and can be neglected. Importantly, these dust rings do not occur at gas pressure maxima, and thus are different from classical dust traps. \citet{LongariniEtal2021} also perform 3D SPH simulations which produce the predicted dust rings, in agreement with the analytical arguments.

In this paper we aim to provide a thorough investigation of the dust pile-ups that occur in inclined circumbinary discs, focusing on the morphology, time evolution, and conditions that enable the formation of these dust density enhancements. We first present our numerical investigations using 3D SPH simulations in Section~\ref{sec:SPH} where we focus on the time evolution and morphology of the resulting dust rings and their dependence on the disc initial inclination and binary eccentricity. We then derive the analytical equations governing the angular momentum transport of dust as it experiences a drag torque from the gas in the general case of a tilted and twisted disc in Section~\ref{sec:analytics}. We implement these equations in a 1D disc evolution code where both gas and dust evolve and interact. This allows us to gain a deeper understanding of the physics underlying these dust pile-ups, as well as perform a parameter sweep of dust grain sizes which we present in Section~\ref{sec:ringcode}. Finally, we discuss our results and summarise the theoretical elements for the formation of these dust pile-ups in Section~\ref{sec:discuss} and conclude in Section~\ref{sec:conclusion}

\section{Smoothed Particle Hydrodynamics Simulations}
\label{sec:SPH}
\citet{Aly&Lodato2020} performed a suite of SPH simulations with 5 different dust species representing different grain sizes to investigate the effects of varying degrees of coupling between gas and dust. Here we take a different approach: we focus only on one dust species with St~$\sim50$ in the weakly coupled regime, which amplifies the effects leading to the dust pile-ups, and use our available computational resources to vary the disc initial inclination $\beta_0$ and the binary eccentricity $e_\mathrm{b}$. This allows us to investigate the dependence of the dust pile-ups on disc and binary parameters. We choose binary eccentricities that span the planar and polar alignment regimes. Due to limited computational resources, we run the majority of our simulations for 1200 binary orbits, which is much longer than the formation timescale of the dust pile-ups. However, we evolve our fiducial setup (circular binary, $\beta_0=40^\circ$) for much longer than the other cases and those reported in \citet{Aly&Lodato2020} to be able to study the long term stability of these pile-ups. We revisit our choice of dust size in Section~\ref{sec:ringcode}. Our simulations parameters are summarized in Table~\ref{table:parameters}.

A warped disc is often characterised by a tilt $\beta$ and twist $\gamma$ angles. The unit angular momentum vector $\hat{\bm l}$ at a certain radius $R$ in the disc is a function of the tilt and twist angles such as \citep{Pringle1996}:
\begin{equation}
    \hat{\bm l} = (\cos\gamma\sin\beta,\sin\gamma\sin\beta,\cos\beta).
\end{equation}
We retain this notation throughout the paper, which is consistent with \citet{LongariniEtal2021}. In general, these angles can vary with $R$. The disc is warped when $\beta$ changes with $R$, and twisted when $\gamma$ changes with $R$. In our simulations, the initial setup is such that the binary is in the X-Y plane, and the disc line of nodes is the Y axis, therefore the tilt angle $\beta$ is simply the angle between the disc angular momentum at any radius and the Z axis.
\begin{table}
\begin{tabular}{ |c|c|r| } 
 \hline
  $\beta_0$ ($^\circ$) & $e_\mathrm{b}$  & End Time (in binary orbits) \\ 
 \hline
 0 & 0 & 1200 \\ 
 20 & 0 & 1200 \\ 
 40 & 0 & (fiducial) 3500 \\
 60 & 0 & 1200 \\
 40 & 0.3 & 1200 \\
 40 & 0.6 & 1200 \\
 \hline
\end{tabular}
\caption{Parameters set for the simulations suite: $\beta_0$ is the disc initial inclination, and $e_\mathrm{b}$ is the binary eccentricity}.
\label{table:parameters}
\end{table}

\subsection{Setup}
We use the SPH code \textsc{phantom} \citep{PriceEtal2018} in all our 3D hydrodynamics simulations. Since we are only simulating large dust particles with St $\sim 50$, we employ the two-fluid algorithm which solves a separate set of evolution equations for the gas and dust, coupled with a drag term. The details of the model can be found in \citet{Laibe&Price2012}, \citet{PriceEtal2018}, and \citet{Price&Laibe2020}. We setup a disc of gas and dust particles around a binary star system with equal mass $M_1=M_2=0.5M_\odot$. The binary stars are represented by sink particles with an accretion radius $R_{\rm acc}=30$ au and a binary semi-major axis $a=50$ au. This large accretion radius ensures that we do not form any circum-stellar discs in our simulations, which reduces the modelled dynamical range and allows us to run our simulations for a longer time. The discs extend from $R_{\rm in}=100$ au to $R_{\rm out}=1000$ au with a total mass $M_{\rm{d}}=10^{-3}M_\odot$. The gas component is represented by $9\times 10^5$ SPH particles and the dust component by $1\times 10^5$ SPH particles. 

The gas particles are distributed such that the initial surface density profile has the power law:
\begin{equation}
    \Sigma_\mathrm{g}=\Sigma_{\mathrm{g},0}\left(\frac{R}{R_{\rm in}}\right)^{-p},
\label{equ:sigma_initial}
\end{equation}
where $\Sigma_{\mathrm{g},0}$ is a normalisation constant. We choose a locally isothermal equation of state with a sound speed $c_\mathrm{s}$ such that:
\begin{equation}
    c_\mathrm{s} = c_\mathrm{s,in} \left(\frac{R}{R_{\rm in}} \right)^{-q}.
\end{equation}
where $c_\mathrm{s,in}$ is the sound speed at the inner radius. We choose power law indices $p=1$ and $q=0.5$. We use a constant SPH artificial viscosity coefficient $\alpha_{\rm SPH}\approx 0.55$ such that the corresponding disc viscosity coefficient \citep{ShakuraSunyaev1973} is $\alpha_\mathrm{d}=0.01$ at the inner radius. We note, however, that our choice of the indices $p$ and $q$ impact the radial dependence of the viscosity coefficient $\alpha_\mathrm{d}$. The mapping between the SPH and disc viscosity coefficient is (e.g., \citealt{Artymowicz&Lubow1994}, \citealt{Murray1996}, \citealt{Lodato&Pringle2007}):
\begin{equation}
    \alpha_\mathrm{d} \propto \alpha_{\rm SPH} \frac{\langle h \rangle}{H}
\end{equation}
where $\langle h \rangle$ is the vertically averaged smoothing length and $H$ is the disc thickness, which depends on disc radius:
\begin{equation}
    H = \frac{c_\mathrm{s}}{\Omega_\mathrm{k}} \propto R^{3/2-q}
\end{equation}
The average smoothing length in 3D has the radial dependence:
\begin{equation}
    \langle h \rangle \propto \rho^{-1/3} \propto \left(\frac{\Sigma}{H}\right)^{-1/3} \propto R^{1/2+(p-q)/3}
\end{equation}
Hence, the choice of $p$ and $q$ determines the radial dependence of the disc viscosity coefficient. A common choice is to use $p=3/2$ and $q=3/4$, which results in a radially constant $\alpha_\mathrm{d}$ (e.g., \citealt{LodatoPrice2010}). In this paper we decide to use a physically motivated choice of indices (e.g., \citealt{AlyEtal2018}), even though it results in a non-constant disc viscosity coefficient. For our choice of $p=1$ and $q=0.5$, the resulting radial dependence of the disc viscosity coefficient is $\alpha_\mathrm{d} \propto R^{-1/3}$. To prevent particle interpenetration, we employ the recommended value $\beta_\mathrm{SPH}=2$ for the quadratic viscosity coefficient \citep{Price&Federrath2010,Meru&Bate2012}.

The particles are distributed vertically according to a Gaussian distribution, as dictated by hydrostatic equilibrium, which results in an aspect ratio profile:
\begin{equation}
    \frac{H}{R} = \frac{c_\mathrm{s}}{\Omega_\mathrm{k} R}=\left( \frac{H}{R} \right)_{\rm in} \left( \frac{R}{R_{\rm in}} \right) ^{1/2-q}
\end{equation}
and we use $\left( \frac{H}{R} \right)_{\rm in}=0.1$. This setup ensures that the disc vertical thickness is resolved by $\sim 5-7$ smoothing lengths at most radii throughout the simulations.

Dust particles are originally distributed following the gas disc profiles, albeit with a dust-to-gas mass ratio of 0.01. We use a uniform dust particles size $s=200 \mu$m and intrinsic density $\rho_\mathrm{d}=5$g/cm$^3$, which for our choice of parameters result in St~$\sim50$. The relatively large St for this grain size is due to the large disc extension which reduces the gas surface density. The disc initial inclination and binary eccentricity are varied per simulation according to Table~\ref{table:parameters}. We note that in this paper we do not make explicit connections with observations at specific wavelengths, and hence the dynamics are characterized by St and the problem is scalable. Therefore, in the remainder of the paper we do not make explicit references to disc radii in au, instead we parameterise locations in the disc by the ratio $R/a$.

\subsection{Dust Traffic Jam}
\begin{figure*}
  \begin{center}
    \resizebox{170.0mm}{!}{\mbox{\includegraphics[angle=0]{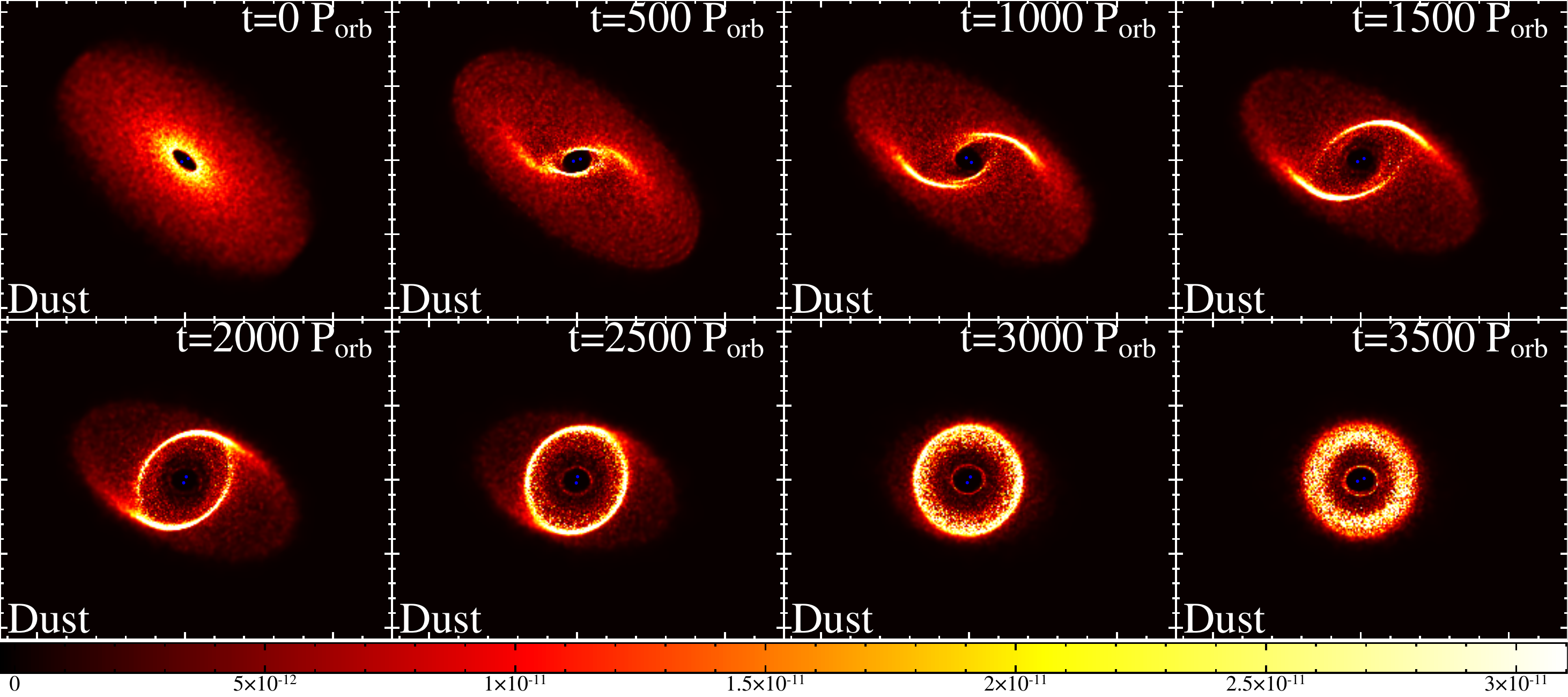}}}
 \caption{\label{fig:ficucial_dens_dust}
 Dust column density in $M_\odot/\mathrm{au}^2$ at different times (in binary orbits) for our fiducial run. The projection plane is fixed in all snapshots.}
 \end{center}
\end{figure*}

\begin{figure*}
  \begin{center}
    \resizebox{170.0mm}{!}{\mbox{\includegraphics[angle=0]{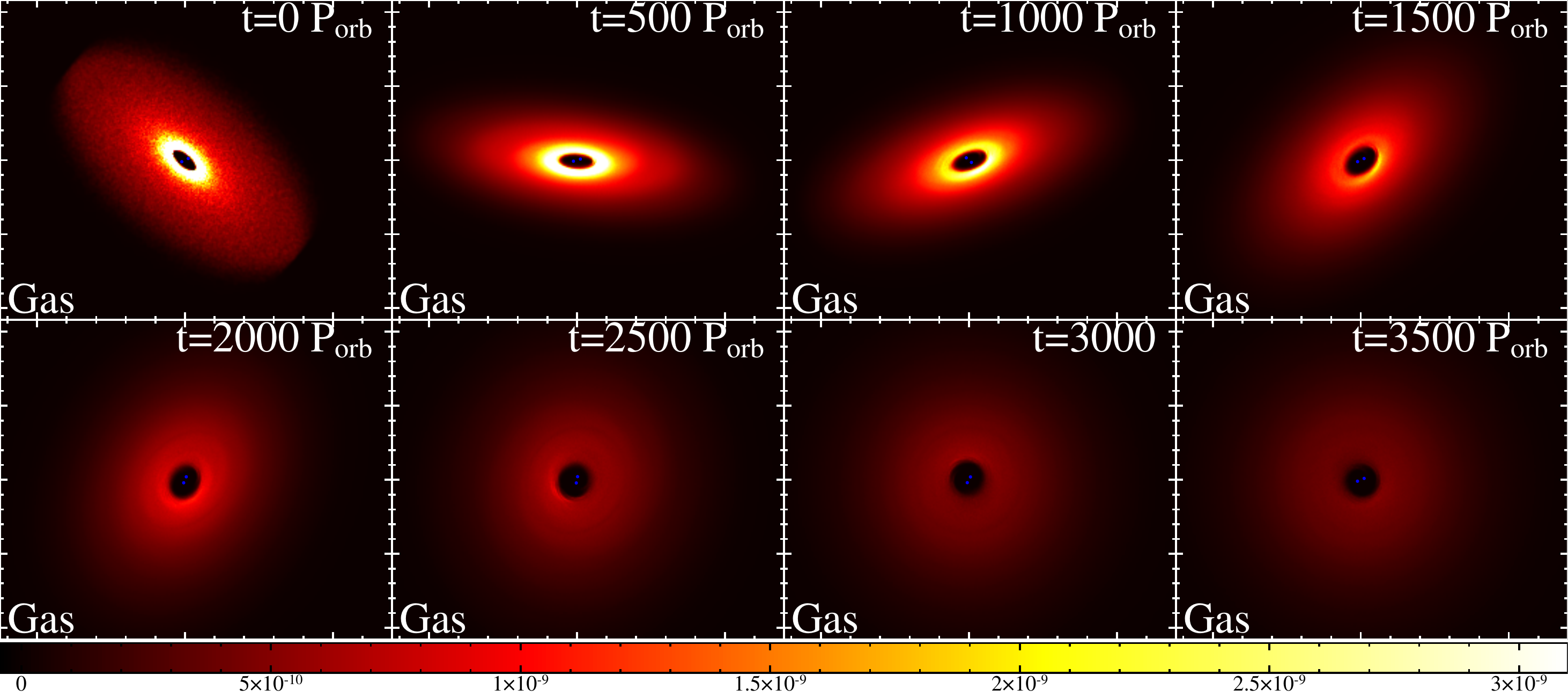}}}
 \caption{\label{fig:ficucial_dens_gas}
 Same as Figure~\ref{fig:ficucial_dens_dust} but for gas.}
 \end{center}
\end{figure*}

Figure~\ref{fig:ficucial_dens_dust} shows dust column density maps at various times for our fiducial simulation of a disc initially tilted by $40^\circ$ around a circular binary. We rotate the view by $30^\circ$ around the Z-axis and $60^\circ$ around the X-axis for better visualisation of the warped and twisted disc. Times denoted in the panels are in units of binary orbits (for reference, $1000$ binary orbits $\sim 3.5 \times 10^5$ years). We note the formation of a prominent dust density enhancement that initially takes the shape of two wide spirals/arcs that gradually gets wound up until it forms a ring after $2000$ binary orbits and persists till the end of the simulation at $3500$ binary orbits ($\sim 1.2 \times 10^6$ years). We refer to this feature as a dust traffic jam in this paper (we provide a justification for the terminology in Section~\ref{sec:discuss}). Figure~\ref{fig:ficucial_dens_gas} shows the gas density columns for the same simulation, times, and viewing angles. Comparing the two figures, one may observe that the gas precesses in a rigid manner, as expected, and retains almost constant twist and tilt angles at any given time, while the dust shows signs of radially dependent twist and tilt profiles. We also see that the gas disc is smooth compared to the dust disc; there are no gas features that correspond to the dust structure in the traffic jams. However, careful examination shows that the traffic jam happens close to the radius where the dust disc orientation coincides with that of the gas disc. Finally, we note from Figure~\ref{fig:ficucial_dens_dust} that after about $2500$ binary orbits ($\sim 0.88 \times 10^6$ years) the dust disc is significantly truncated and material outside the traffic jam feature has largely drifted inwards.

\begin{figure}
\resizebox{85.0mm}{!}{\mbox{\includegraphics[angle=0]{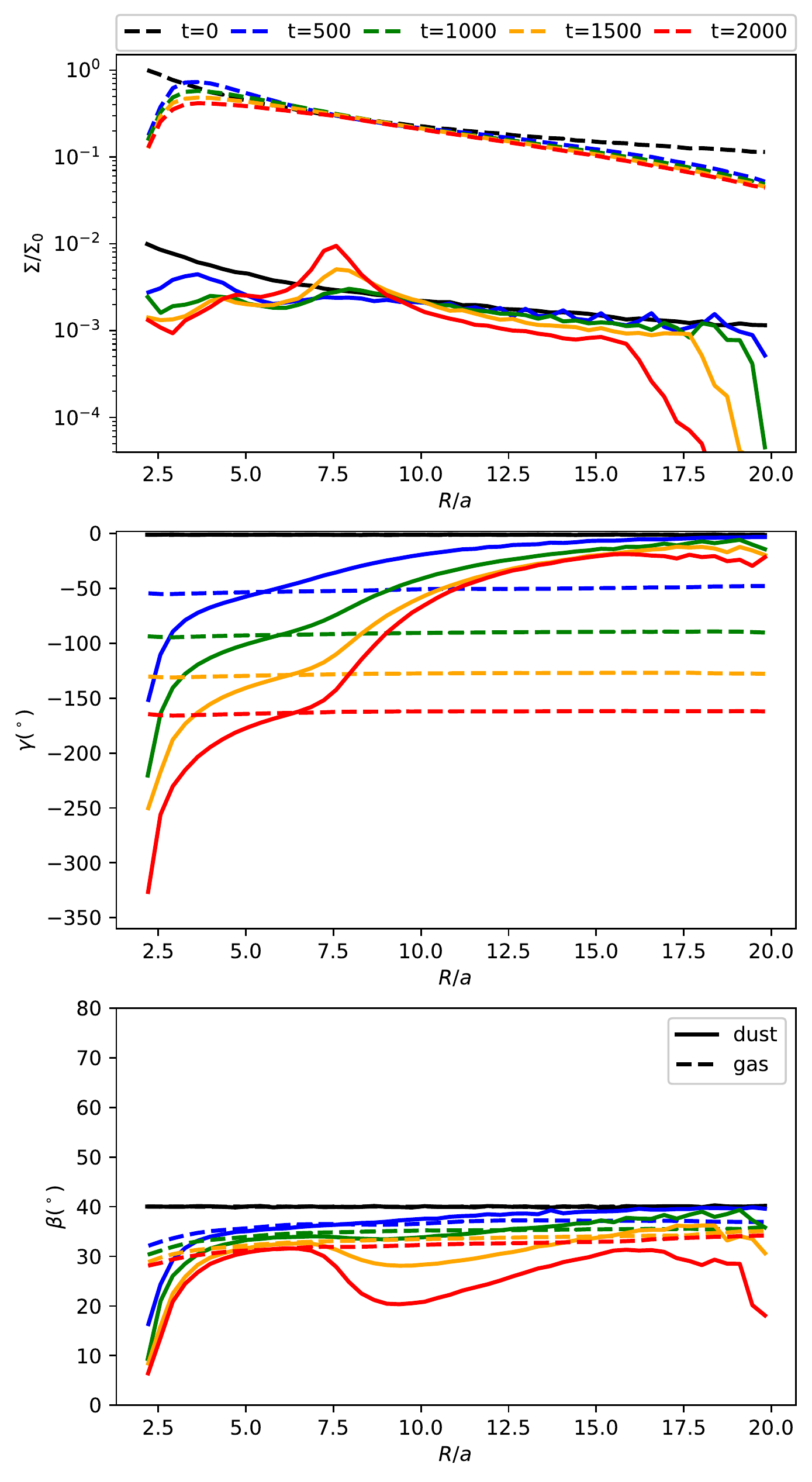}}}

\caption{\label{fig:fiducial_profile}
Radial profiles of surface density (top), twist angle (middle), and tilt angle (bottom) for dust (solid lines) and gas (dashed lines) in the fiducial SPH simulation. Times are in multiples of binary orbits.}
\end{figure} 

Figure~\ref{fig:fiducial_profile} shows the azimuthally-averaged radial surface density (top panel), twist (middle panel), and tilt (bottom panel) angles profiles for our fiducial run for both the gas and dust components at several times. We notice a bump in the dust surface density at $R/a\sim 7.5$ which corresponds to the traffic jam shown in Figure~\ref{fig:ficucial_dens_dust}. No corresponding feature exists for the gas at this radius, which is evidence that this is different from classical dust traps that occur at gas pressure maxima. We also notice the enhanced clearing of the dust disc at outer radii. The twist angle profiles in the middle panel show that the gas precesses rigidly and retains an almost constant twist angle throughout the disc, while dust exhibits a radial dependence akin to that of test particles (see Figures~\ref{fig:dust_surface_render} and~\ref{fig:gas_surface_render} for a clear display of the warps). Importantly, we note that in all four times shown in the figure, the gas and dust twist profiles intersect at $R/a\sim 6$, at a slightly smaller radius than the location of the dust density bump. This is indeed the co-precession radius reported in \citet{Aly&Lodato2020} and \citet{LongariniEtal2021} (indeed this is in perfect agreement with the radius obtained from equation~\ref{equ:Rcp} derived in \citet{LongariniEtal2021}). The bottom panel shows the tilt profiles for both phases. We notice that the gas slowly and uniformly decays towards planar alignment with the binary; i.e, $\beta=0$, as a result of viscous dissipation. The dust on the other hand displays a more involved behaviour, which we fully explain in Section~\ref{sec:discuss}. 

Dust in a flat aligned disc undergoes radial drift because of the velocity difference between gas and dust at the same radius due to the reduction of gas orbital speed caused by the pressure support. This reduction is typically of the order of $(c_\mathrm{s}/v_\mathrm{k})^2 \sim (H/R)^2$ (where $v_\mathrm{k}$ is the Keplerian velocity). Now consider the situation in our fiducial simulation: the difference in twist angle between the dust and gas increases with time. Indeed we can see from Figure~\ref{fig:fiducial_profile} that at $t=2000$ binary orbits, the difference in phase approaches $150^\circ$ in the outer disc, where the dust is hardly precessing at all. This greatly increases the headwind that the dust feels in the outer disc, compared to the planar case. The dust drifts radially much more quickly. We also note that the slope of the dust twist profile becomes shallower around the co-precession radius, as a consequence of the increased coupling between gas and dust at that radius where the orbits coincide. The change in the slope of the dust twist profile creates an inflection point at a radius just outside the co-precession radius. As dust particles start drifting from the outer disc at enhanced rates, the twist angle difference driving this drift decreases and dust particles slow their drift inwards, creating a traffic jam where the drift velocity is expected to be minimum.

\subsection{Effects of varying disc inclination}
\begin{figure*}
  \begin{center}
    \resizebox{175.0mm}{!}{\mbox{\includegraphics[angle=0]{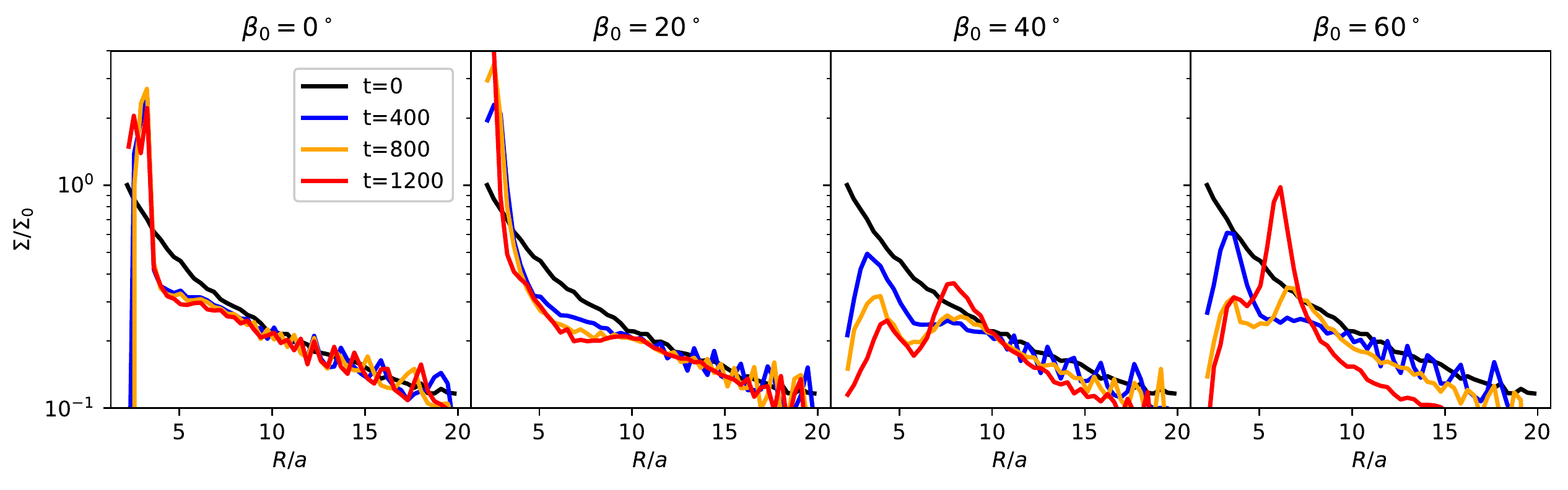}}}
 \caption{\label{fig:beta}
 Dust surface density profiles for 4 simulations with 4 different initial inclinations $\beta_0=0, 20, 40, \rm{and}\ 60^\circ$.}
 \end{center}
\end{figure*}
In Figure~\ref{fig:beta} we plot the dust surface density profiles for four different simulations with varying initial inclination $\beta_0$ (left to right $\beta_0=0^\circ, 20^\circ, 40^\circ, 60^\circ$ at different times. The strength of the traffic jam increases with initial inclination, as the effect of the binary gravitational torque causing the traffic jam increases (see section~\ref{sec:analytics}). This is accompanied by an accelerated radial drift at higher initial disc inclination. The two phenomena are intrinsically linked; the traffic is driven by an accelerated radial drift everywhere in the disc, which is minimised near the co-precession radius where the traffic jam occurs. At low initial disc inclination, we notice a spike in dust surface density at the inner edge of the disc. This spike is at least partly numerical due to the reduction of the gas resolution at the inner edge due to binary tidal effects, which may cause dust clumping below the gas resolution scale \citep{PriceEtal2018}.

A careful examination of the time evolution of the surface density for the $\beta_0=60^\circ$ and $\beta_0=40^\circ$ cases in Figure~\ref{fig:beta} shows that two density enhancements form: an inner density enhancement at $R/a\sim 3.5$ forms first, and then slowly decreases as a more prominent bump forms further out at $R/a\sim 7$. Figure~\ref{fig:fiducial_profile} shows that the co-precession radius for the $\beta_0=40^\circ$ case is between these two radii. This is in line with the prediction of two dust pile-ups in \citep{LongariniEtal2021}. Since the difference in twist angle is what drives the traffic jam, we expect the inner one to form first since precession is faster in the inner disc. At later times, however, the outer traffic jam starts forming and starves the inner one of material, since dust drifting from the outer disc meets the outer ring first. This behaviour can also be seen in Figure 1 in \citet{Aly&Lodato2020}. Even though they focus on the inner dust pile-up, we can see the outer one starts to develop at the last snapshot, and would have overtaken the inner bump had they run their simulations for longer.

\subsection{Effects of varying binary eccentricity}

\begin{figure*}
  \begin{center}
    \resizebox{175.0mm}{!}{\mbox{\includegraphics[angle=0]{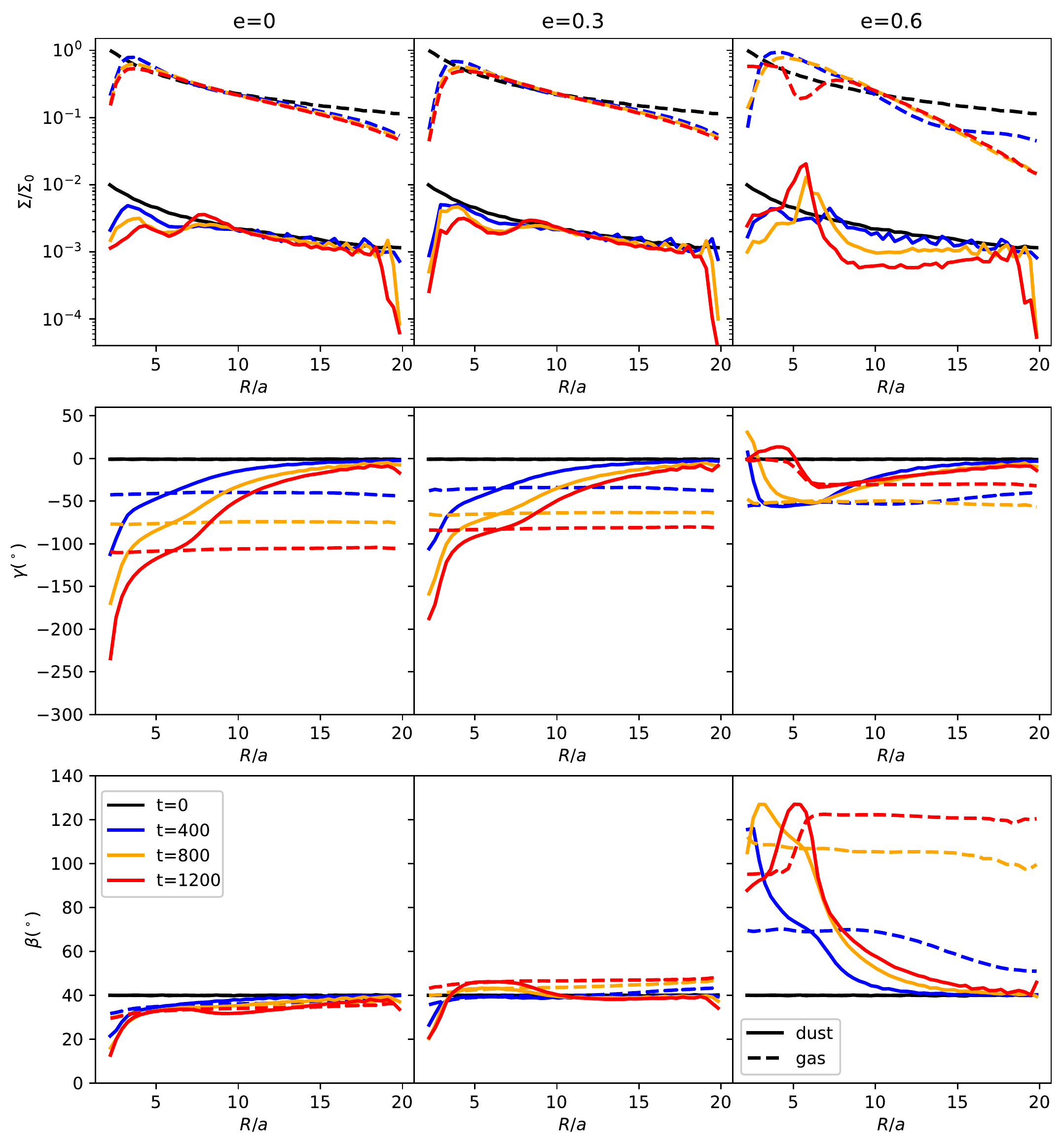}}}
 \caption{\label{fig:ecc_profile}
 Same as Figure~\ref{fig:fiducial_profile} but for 3 simulations with 3 different binary eccentricities $e_b=0.0, 0.3, \rm{and}\ 0.6$.}
 \end{center}
\end{figure*}
In Figure~\ref{fig:ecc_profile} we plot the radial profiles for the surface density, twist, and tilt angles for cases with binary eccentricities $e_b=0.0, 0.3, \rm{and}\ 0.6$ all with the same disc initial inclination $\beta_0=40^\circ$. We note that $e_b=0.6$ is expected to be in the polar alignment regime (since the critical tilt angle in this case is $31.75^\circ$, as computed from \citet{Martin&Lubow2019}), while the other two cases are expected to be in the planar alignment regime. For the general case of a test particle orbiting an eccentric binary, equation~\ref{equ:precess_circular} becomes (e.g., \citealt{AlyEtal2015}):
\begin{equation}
   \bm{\Omega}_\mathrm{p}{(R)}=\frac{3}{4} \frac{\sqrt{G M} \eta a^{2}}{R^{7 / 2}}[5e_\mathrm{b}^2(\hat{\bm{l}}\cdot \hat{\bm{e}})\hat{\bm{e}}-(1-e_\mathrm{b}^2)(\hat{\bm{l}}\cdot \hat{\bm{h}})\hat{\bm{h}}]
    \label{equ:precess_ecc}
\end{equation}
where $\bm{e}$ is the binary eccentricity vector and $\bm{h}$ is the binary specific angular momentum vector. Equation~\ref{equ:precess_ecc} implies that for the $e_\mathrm{b}=0.3$, planar precession dominates but with a precession rate lower than the circular binary case, whereas for the $e_\mathrm{b}=0.6$ case, polar precession is the dominant mode, with a precession rate higher than both the circular binary and the $e_\mathrm{b}=0.3$ cases. This is exactly what we observe in Figure~\ref{fig:ecc_profile}. The twist angle profile for the $e_\mathrm{b}=0.3$ case resembles that of the circular binary case, albeit with slower evolution. This results in a less prominent traffic jam as shown in the surface density profile, as expected. The tilt angle profile for the $e_\mathrm{b}=0.6$ case clearly shows that the disc is precessing towards polar alignment around the binary. In fact, we can notice that the tilt and twist angles profiles switch trends compared with the $e_\mathrm{b}=0.3$ case (for example, notice that the tilt profile for the $e_\mathrm{b}=0.6$ case resembles a mirrored version of the twist profile of the circular or $e_\mathrm{b}=0.3$ cases), as expected since in the polar alignment regime precession around the vector $\bm{e}$ dominates, rather than around the vector $\bm{h}$. Therefore, the location of the traffic jam, seen in the surface density profile, is better predicted by the radius at which the gas and dust tilt angles coincide, rather than their twist angles in the planar alignment cases. 

The stronger precession torque in the $e_\mathrm{b}=0.6$ case (as can be inferred from equation~\ref{equ:precess_ecc}) has two consequences: first, the faster precession leads to quicker detachment between the gas and disc angles, resulting in a faster radial drift at the outer edge and a significantly more prominent traffic jam, as is evident in the surface density profile. Second, the gas disc breaks at a radius close to that of the dust traffic jam, as can be seen in the gas tilt and surface density profiles at $R/a\sim 6$. However, the time evolution shown in Figure~\ref{fig:ecc_profile} shows that the dust traffic jam had already developed by $t=800$ binary orbits, while the disc only shows evidence of breaking at $t=1200$ binary orbits, indicating that the disc break is not the cause of the traffic jam. The disc break somewhat complicates the phenomenon: instead of one gas disc precessing rigidly, we now have two gas discs with different precession frequencies. In the case simulated here, this difference in precession hampers the formation of the inner density enhancement. We note, however, that the prominent outer traffic jam has already evolved to a near-polar configuration, possibly allowing for the formation of polar planetesimals. In general, the interplay between disc breaking and traffic jams formation depends on the radius and evolution stage at which the disc breaks, which in turn depends on the disc and binary parameters. The general outcome of such interaction cannot be generalised from this particular simulation here and will need to be investigated systematically in future studies.

\begin{figure*}
  \begin{center}
    \resizebox{170.0mm}{!}{\mbox{\includegraphics[angle=0]{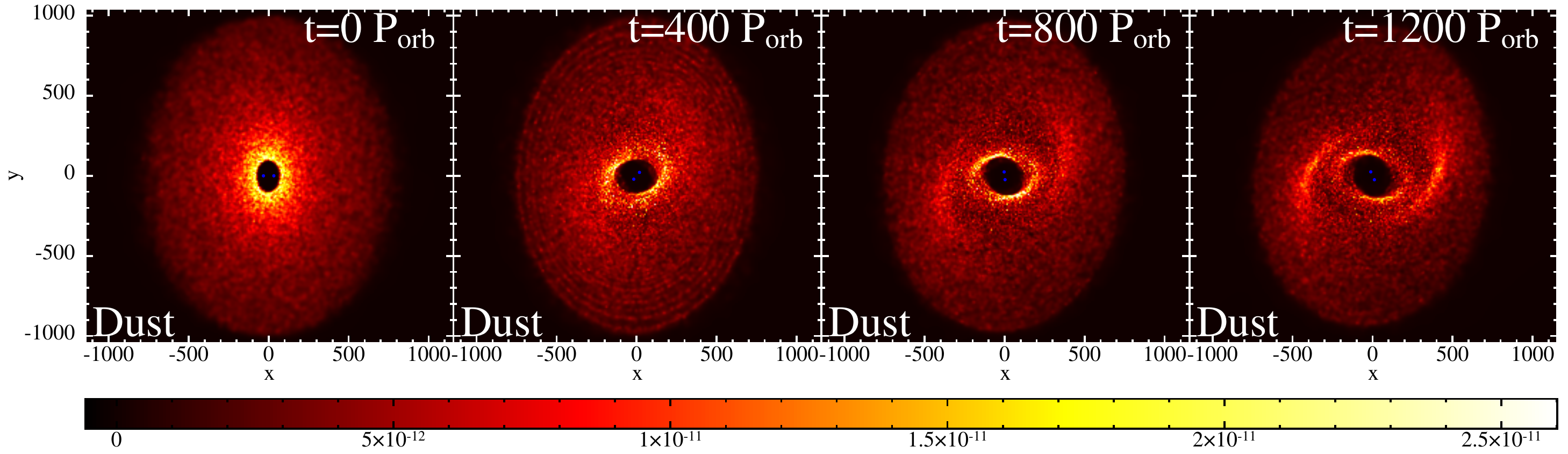}}}
    \resizebox{170.0mm}{!}{\mbox{\includegraphics[angle=0]{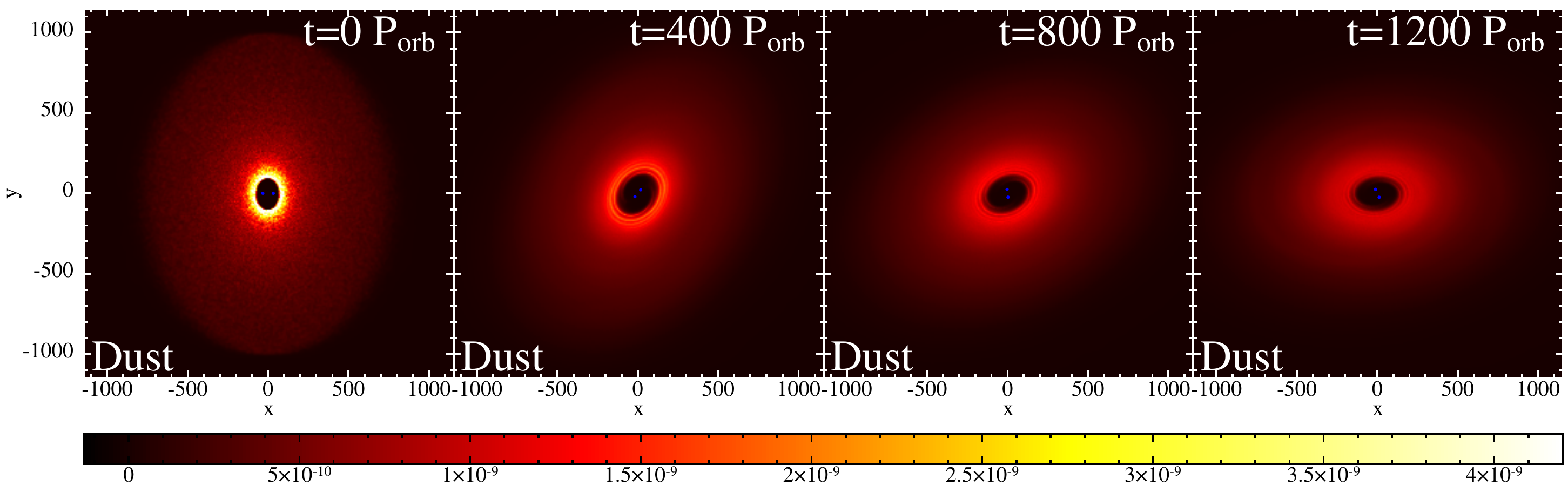}}}

 \caption{\label{fig:ecc0.3_dens}
 Dust and gas column density projected onto the X-Y plane at 4 different times (indicated in units of binary orbits) for the simulation with binary eccentricity $e_\mathrm{b}=0.3$. }
 \end{center}
\end{figure*}

\begin{figure*}
  \begin{center}
    \resizebox{170.0mm}{!}{\mbox{\includegraphics[angle=0]{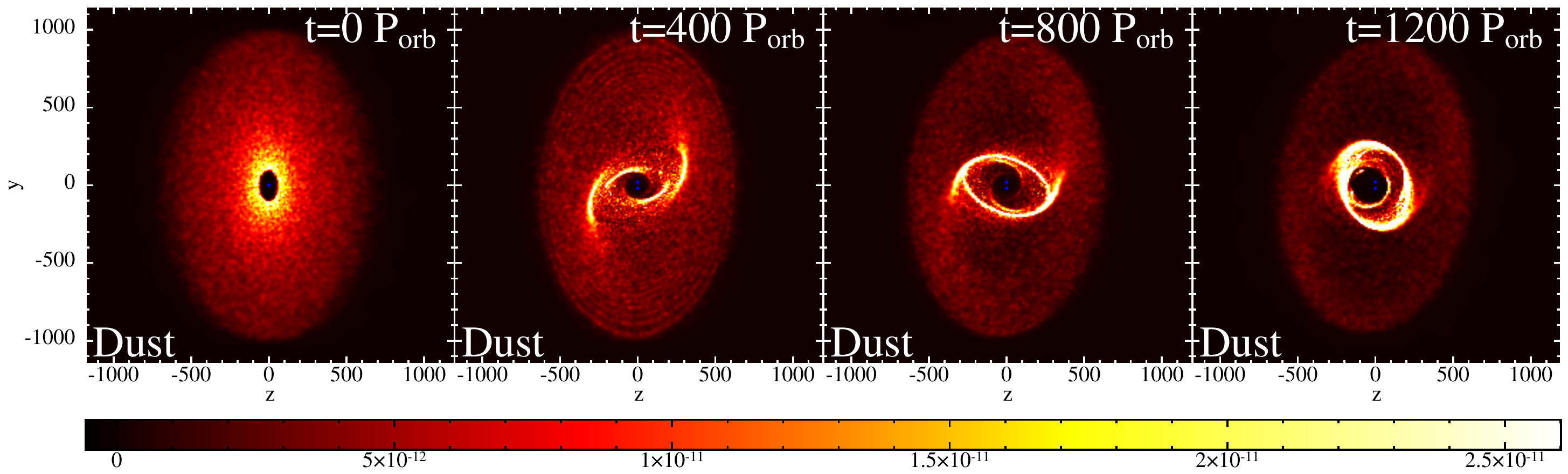}}}
    \resizebox{170.0mm}{!}{\mbox{\includegraphics[angle=0]{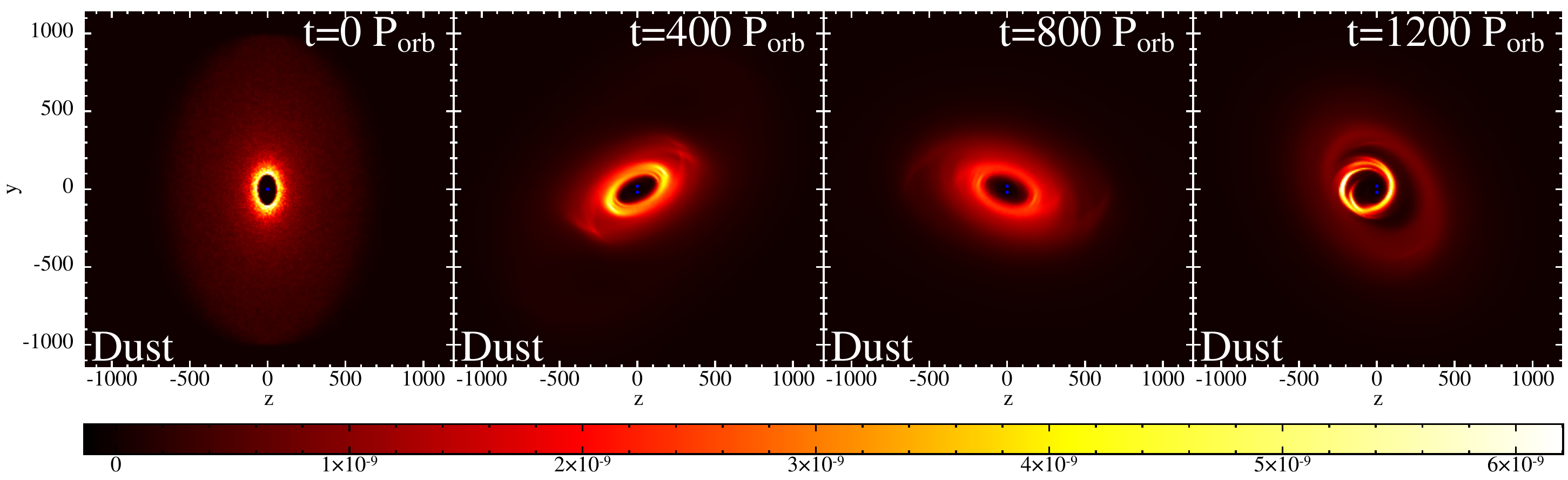}}}

 \caption{\label{fig:ecc0.6_dens}
 Same as Figure~\ref{fig:ecc0.3_dens} but for $e_\mathrm{b}=0.6$ and projected onto the Y-Z plane.}
 \end{center}
\end{figure*}

Column density of gas and dust of the $e_\mathrm{b}=0.3$ and $e_\mathrm{b}=0.6$ cases are shown at different times in Figures~\ref{fig:ecc0.3_dens} and~\ref{fig:ecc0.6_dens}; respectively. We chose a projection plane that shows the traffic jam evolution close to a face-on view. The different precession behaviour meant that these planes differ in the planar and polar cases; hence Figure~\ref{fig:ecc0.3_dens} shows an X-Y projection, while Figure~\ref{fig:ecc0.6_dens} shows an Z-Y projection. We can see that the traffic jam is indeed more prominent in the $e_\mathrm{b}=0.6$ case. The gas disc break is also clear in Figure~\ref{fig:ecc0.6_dens}.


\section{A Model For Dust Evolution}
\label{sec:analytics}
Assuming a homogeneous distribution of dust particles with a constant mass and intrinsic density, the gas drag force acting on one single dust particle is:
\begin{equation}
    \bm{F}=k_\mathrm{s}(\bm{v}_\mathrm{g}-\bm{v}_\mathrm{d})
    \label{equ:drag_p}
\end{equation}
with $\bm{v}_\mathrm{g}$ and $\bm{v}_\mathrm{d}$ as the gas and dust velocities, and $k_\mathrm{s}$ is a drag coefficient that depends on the gas mean free path and the dust particle size (see for example \citet{BainesEtal1965} and \citet{Fan&Zhu1998}).

We now consider the force exerted by a ring of gas on a ring of dust. Both rings are at radius $R$ with a width of $\mathrm{d}R$, but they can be inclined with respect to each other by an angle $\theta$ and have thicknesses of $H_\mathrm{g}$ and $H_\mathrm{d}$ for the gas and dust rings, respectively. The intersection volume between the two rings can be approximated by:
\begin{equation}
    V\approx\frac{2 H_\mathrm{g} H_\mathrm{d} \mathrm{d}R}{\sin\theta+\frac{H_\mathrm{g}}{\upi R}}.
\end{equation}
where the second term in the denominator ensures the expression converges to the the volume of the dust ring as the mutual inclination goes to zero. The drag force acting on the dust ring $F_\mathrm{r}$ is the drag force per particle (equation~\ref{equ:drag_p}) multiplied by the number of particles in the intersection volume:
\begin{equation}
    \bm{F}_\mathrm{r}=\frac{2\rho_\mathrm{d} k_\mathrm{s} H_\mathrm{g} H_\mathrm{d} \mathrm{d}R}{m_\mathrm{s}\left(\sin\theta+\frac{H_\mathrm{g}}{\upi R}\right)}(\bm{v}_\mathrm{g}-\bm{v}_\mathrm{d}),
\end{equation}
where $m_\mathrm{s}$ is the dust particle mass and $\rho_\mathrm{d}$ is the dust volume density in the ring . Assuming that vertical variation in $\rho_\mathrm{d}$ can be neglected; i.e $\Sigma_\mathrm{d}=\rho_\mathrm{d} H_\mathrm{d}$, the above equation becomes:
\begin{equation}
    \bm{F}_\mathrm{r}=\frac{2k_\mathrm{s} H_\mathrm{g} \mathrm{d}R}{m_\mathrm{s}\left(\sin\theta+\frac{H_\mathrm{g}}{\upi R}\right)}(\Sigma_\mathrm{d}\bm{v}_\mathrm{g}-\Sigma_\mathrm{d}\bm{v}_\mathrm{d}).
\end{equation}
The torque density on the dust ring due to the drag from the gas ring is thus:
\begin{equation}
    \bm{T}_\mathrm{d}=\frac{\bm{R}\times\bm{F}_\mathrm{r}}{2\upi R \mathrm{d}R}= \frac{k_\mathrm{s}H_\mathrm{g} (\epsilon\bm{L}_\mathrm{g}-\bm{L}_\mathrm{d})}{m_\mathrm{s}(R\upi \sin\theta + H_\mathrm{g})}
\label{equ:torque}
\end{equation}
where $\epsilon=\Sigma_\mathrm{d}/\Sigma_\mathrm{g}$ is the dust to gas ratio, and $\bm{L}_\mathrm{g}$ and $\bm{L}_\mathrm{d}$ are the gas and dust angular momentum densities
\begin{equation}
    \bm{L}_\mathrm{g}=\Sigma_\mathrm{g} R^2 \Omega_\mathrm{g} \bm{l}_\mathrm{g}
\end{equation}
and
\begin{equation}
    \bm{L}_\mathrm{d}=\Sigma_\mathrm{d} R^2 \Omega_\mathrm{d} \bm{l}_\mathrm{d}
\end{equation}
With $\Omega_\mathrm{g}$ and $\Omega_\mathrm{d}$ representing the gas and dust orbital frequencies; respectively. They are assumed to be close to, but not necessarily equal to the Keplerian frequency.

Following \citet{Papaloizou&Pringle1983}, and treating dust as purely inviscid, the mass conservation equation for the dust is
\begin{equation}
\frac{\partial \Sigma_\mathrm{d}}{\partial t}+\frac{1}{R}\frac{\partial}{\partial R}(R\Sigma_\mathrm{d} V_{R\mathrm{d}})=0
\label{mass}
\end{equation}
where $V_{R\mathrm{d}}$ is the dust radial drift velocity. The angular momentum conservation for the dust is
\begin{equation} 
\frac{\partial \bm{L}_\mathrm{d}}{\partial t }+\frac{1}{R}\frac{\partial }{\partial R}(\Sigma_\mathrm{d} V_{R\mathrm{d}} R^3 \Omega_\mathrm{d} \bm{l}_\mathrm{d})=\bm{T}_\mathrm{d}.
\label{angmom}
\end{equation}
By manipulating equations~\ref{mass} and~\ref{angmom}, (namely; we multiply equation~\ref{mass} by $R^2 \Omega_\mathrm{d}$ and substract the result from the dot product of equation~\ref{angmom} with $\bm{l}_\mathrm{d}$), we obtain an expression for the dust radial drift velocity $V_{R\mathrm{d}}$
\begin{equation}
V_{R\mathrm{d}}=\frac{\bm{T}_\mathrm{d}\cdot \bm{l}_\mathrm{d}}{\Sigma_\mathrm{d} \, \mathrm{d}(R^2 \Omega_\mathrm{d})/ \mathrm{d}R}
\label{vr}
\end{equation}
and the dust angular momentum evolution equation becomes:
\begin{equation}
\frac{\partial \bm{L}_\mathrm{d}}{\partial t}  = \bm{T}_\mathrm{d}
 -\frac{1}{R} \frac{\partial }{\partial R}\left[ \frac{R(\bm{T}_\mathrm{d} \cdot \bm{l}_\mathrm{d} )}{\Sigma_\mathrm{d} \, \mathrm{d} (R^2 \Omega_\mathrm{d})/\mathrm{d} R}\bm{L}_\mathrm{d}\right].
\end{equation}
For a ring orbiting with near-Keplerian velocity, the above 2 expressions become
\begin{equation}
V_{R\mathrm{d}}=\frac{2\bm{T}_\mathrm{d}.\bm{l}_\mathrm{d}}{\Sigma_\mathrm{d} R \Omega_\mathrm{d}}
\label{equ:radial_v_keplerian}
\end{equation}
and
\begin{equation}
\frac{\partial \bm{L}_\mathrm{d}}{\partial t}  = \bm{T}_\mathrm{d}
 -\frac{2}{R} \frac{\partial }{\partial R}\left[ \frac{ R(\bm{T}_\mathrm{d} \cdot \bm{l}_\mathrm{d} )}{\Sigma_\mathrm{d} R \Omega_\mathrm{d}}\bm{L}_\mathrm{d}\right].
\label{equ:angmom_keplerian}
\end{equation}
Finally, since in this paper we treat circumbinary discs subject to a binary gravitational torque, we add an external torque to the angular momentum equation
\begin{equation}
\frac{\partial \bm{L}_\mathrm{d}}{\partial t}  = \bm{T}_\mathrm{d}
-\frac{2}{R} \frac{\partial }{\partial R}\left[ \frac{ R(\bm{T}_\mathrm{d} \cdot \bm{l}_\mathrm{d} )}{\Sigma_\mathrm{d} R \Omega}\bm{L}_\mathrm{d}\right] + \bm{\Omega}_\mathrm{p}\times \bm{L}_\mathrm{d}
\label{equ:angmom_keplerian_binary}
\end{equation}
where $\bm{\Omega}_\mathrm{p}$ is obtained from equation~\ref{equ:precess_ecc} and $\bm{T}_\mathrm{d}$ from equation~\ref{equ:torque}. Equation~\ref{equ:angmom_keplerian_binary} completely governs the evolution of the dust, since by obtaining $\bm{L}_\mathrm{d}$ we also get the surface density $\Sigma_\mathrm{d}=L_\mathrm{d}/(R^2 \Omega)$ 

A sanity check for our new dust model is to consider the simple case of dust and gas discs with no mutual inclinations. The radial force balance for the gas disc, along with an assumption of a local isothermal equation of state, dictates that the gas orbital frequency is
\begin{equation}
    \Omega_\mathrm{g} = \Omega_\mathrm{k} \sqrt{1-\mu}
\label{equ:omega_g}
\end{equation}
where $\mu \sim 2(H/R)^2$ \citep{Weidenschilling1977}. Using this sub-Keplerian gas orbital frequency for $\bm{L}_\mathrm{g}$ in equation~\ref{equ:torque} and substituting in equation~\ref{equ:radial_v_keplerian}, the expression for the dust radial velocity becomes
\begin{equation}
   V_{R\mathrm{d}} \approx -2 R \frac{\Omega_\mathrm{k}-\Omega_\mathrm{g}}{\mathrm{St}},
\end{equation}
which is an expression that can be obtained from the azimuthal force balance of a dust particle in gas disc, in the limit of high St \citep{Takeuchi&Lin2002}. Moreover, if we consider the effect of gas drag on a dust particle, radial force balance leads to an expression for the dust orbital frequency (ignoring the effects of the radial component for the gas velocity
\begin{equation}
    \Omega_\mathrm{d} = \Omega_\mathrm{k} \left(\frac{\mathrm{St}^2\mu}{2(1+\mathrm{St}^2)}+\sqrt{1-\mu}\right)
\label{equ:omega_d}
\end{equation}
using this to compute $\bm{L}_\mathrm{d}$ in equation~\ref{equ:torque}, equation~\ref{equ:radial_v_keplerian} simplifies to
\begin{equation}
    V_{R\mathrm{d}} \approx \frac{-\mu R \Omega_\mathrm{k} }{\mathrm{St}+\mathrm{St}^{-1}},
\end{equation}
which is the usual formula for the radial drift velocity caused by the gas pressure support \citep{Takeuchi&Lin2002}. Thus by using equations~\ref{equ:omega_g} and~\ref{equ:omega_d} to describe the gas and dust orbital frequencies, we make sure that our dust model reproduces the known radial drift behaviour in the limit where the gas and dust rings are co-planar. In Section~\ref{sec:discuss} we will see that this is only secondary to the radial drift caused by the drag torque at high gas-dust inclinations. 
\section{1D Analysis}
\label{sec:ringcode}
The dust model derived in Section~\ref{sec:analytics} requires knowledge of the gas angular momentum profile in order to compute the drag torque via equation~\ref{equ:torque}. Since the emphasis of this work is on protoplanetary discs, which are expected to have $H/R > \alpha_\mathrm{d}$, we adopt the linearised set of equations that describe warp propagation in the bending wave regime \citep{Papaloizou&Lin1995,Lubow&Ogilivie2000,LubowEtal2002}
\begin{equation}
    \Sigma_\mathrm{g} R^2 \Omega_\mathrm{g}  \frac{\partial \bm{l}_\mathrm{g} }{\partial t} =
\frac{1}{R}\frac{\partial \bm{G}}{\partial R}+\bm{\Omega}_\mathrm{p}\times \bm{L}_\mathrm{g}
\label{equ:gas_1}
\end{equation}
and
\begin{equation}
    \frac{\partial \bm{G}}{\partial t}+ \frac{\Omega_\mathrm{g}^2-\kappa^2}{2\Omega_\mathrm{g}} \, \bm{l}_\mathrm{g}\times \bm{G} + \alpha \Omega_\mathrm{g} \bm{G}=\frac{\Sigma_\mathrm{g} H^2 R^3\Omega_\mathrm{g}^3}{4}\frac{\partial \bm{l}_\mathrm{g}}{\partial R},
\label{equ:gas_2}
\end{equation}
where $\bm{G}$ is the internal disc torque, and $\kappa$ is the epicycle frequency. The second term on the left hand side is an apsidal frequency term due to non-keplerianity. The gas surface density is prevented from evolving in time; i.e $\partial \Sigma_\mathrm{g} / \partial t=0$, an assumption which is justified for most of our simulations (e.g., see Figure~\ref{fig:fiducial_profile}).

\subsection{Comparison with SPH}
We use the 1D code \textsc{RiCo} to compare with our 3D results. In the gas only regime, \textsc{RiCo} faithfully reproduces results from \cite{LodatoPrice2010} and \cite{MartinEtal2019}. As in \cite{LodatoPrice2010} we assume a zero torque boundary condition. The evolution of the gas is governed by a first order, finite difference, time-explicit scheme for equations~\ref{equ:gas_1} and~\ref{equ:gas_2} and that of the dust by equations~\ref{equ:torque} and~\ref{equ:angmom_keplerian_binary}. We note that our system of equations does not take the back-reaction of the dust on the gas into account (which is included in the SPH treatment). We discretise the disc into 100 linearly spaced grid cells (increasing the resolution beyond that did not result in a noticeable change in the results).  
We use the same disc and binary parameters presented in Section~\ref{sec:SPH}, so that we are able to compare directly with the 3D SPH results. However, we experiment with varying 2 parameters in our 1D code: The initial gas surface density profile $\Sigma_\mathrm{g}(R)$ and the viscosity coefficient $\alpha_\mathrm{d}$. In figure~\ref{fig:fiducial_profile} we notice that the gas surface density profile in our SPH simulations quickly relaxes from the pure power law initial conditions to a steeper power law with a taper at the inner edge and an exponential cutoff at the outer edge. We approximate this steady state by the profile:
\begin{equation}
    \Sigma_\mathrm{g} = \Sigma_{\mathrm{g},0}(R/R_\mathrm{in})^{-3/2}(1 - (R_\mathrm{in}/R)^{1/2})\exp(-(R/R_\mathrm{out})^{1/2}).
\end{equation}

\begin{figure}
\resizebox{85.0mm}{!}{\mbox{\includegraphics[angle=0]{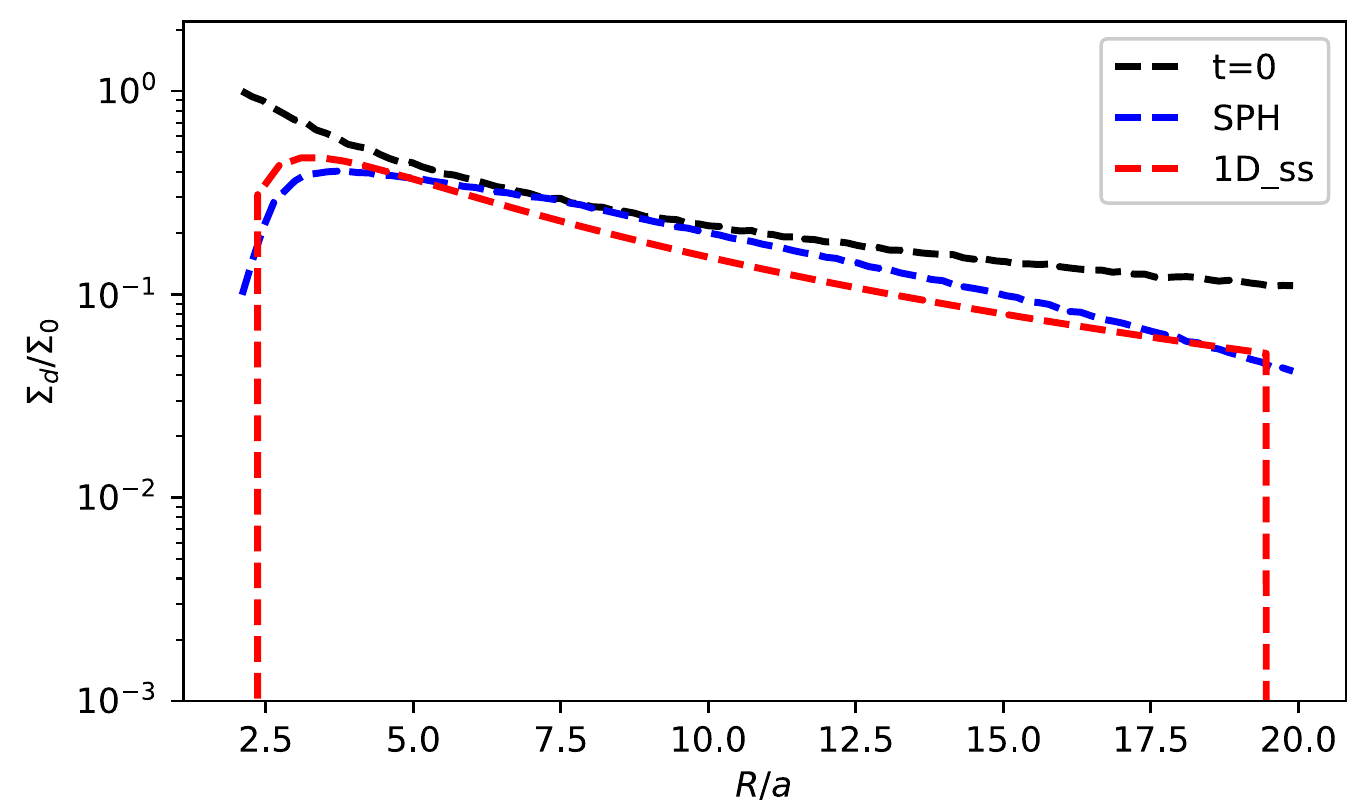}}}

\caption{\label{fig:SPHvRingGas}
Gas surface density profile for the SPH initial conditions (pure power law, black), SPH at t=2000 orbits (blue), and our approximation implemented in the 1D code to represent the SPH steady state denoted $\mathrm{1D}_{ss}$ (red).}
\end{figure}

Since the surface density profile in our 1D code is not allowed to evolve, we use this steady state approximation as well as the pure power law shown in equation~\ref{equ:sigma_initial}. We show these gas surface density profiles in Figure~\ref{fig:SPHvRingGas}. While we do not intend to systematically fit the SPH gas surface density, the above expression is a much better representation of the steady state surface density than a pure power law, especially at the inner and outer boundaries where the binary and drag torques are most significant.  Moreover, we experiment with using a disc viscosity coefficient $\alpha_\mathrm{d}=0.05$ as well as the fiducial $\alpha_\mathrm{d}=0.01$. The increased $\alpha_\mathrm{d}$ value is meant to test the effects of possible SPH excess dissipation which can be due the contribution of the $\beta_\mathrm{SPH}$ term \citep{Meru&Bate2012}, the bulk viscosity that inevitably results from mapping SPH artificial viscosity into physical viscosity \citep{LodatoPrice2010}, or particle noise and re-ordering especially in the case of strong shear flow (for example, \citealt{DehnenAly2012}).

\begin{figure}
\resizebox{85.0mm}{!}{\mbox{\includegraphics[angle=0]{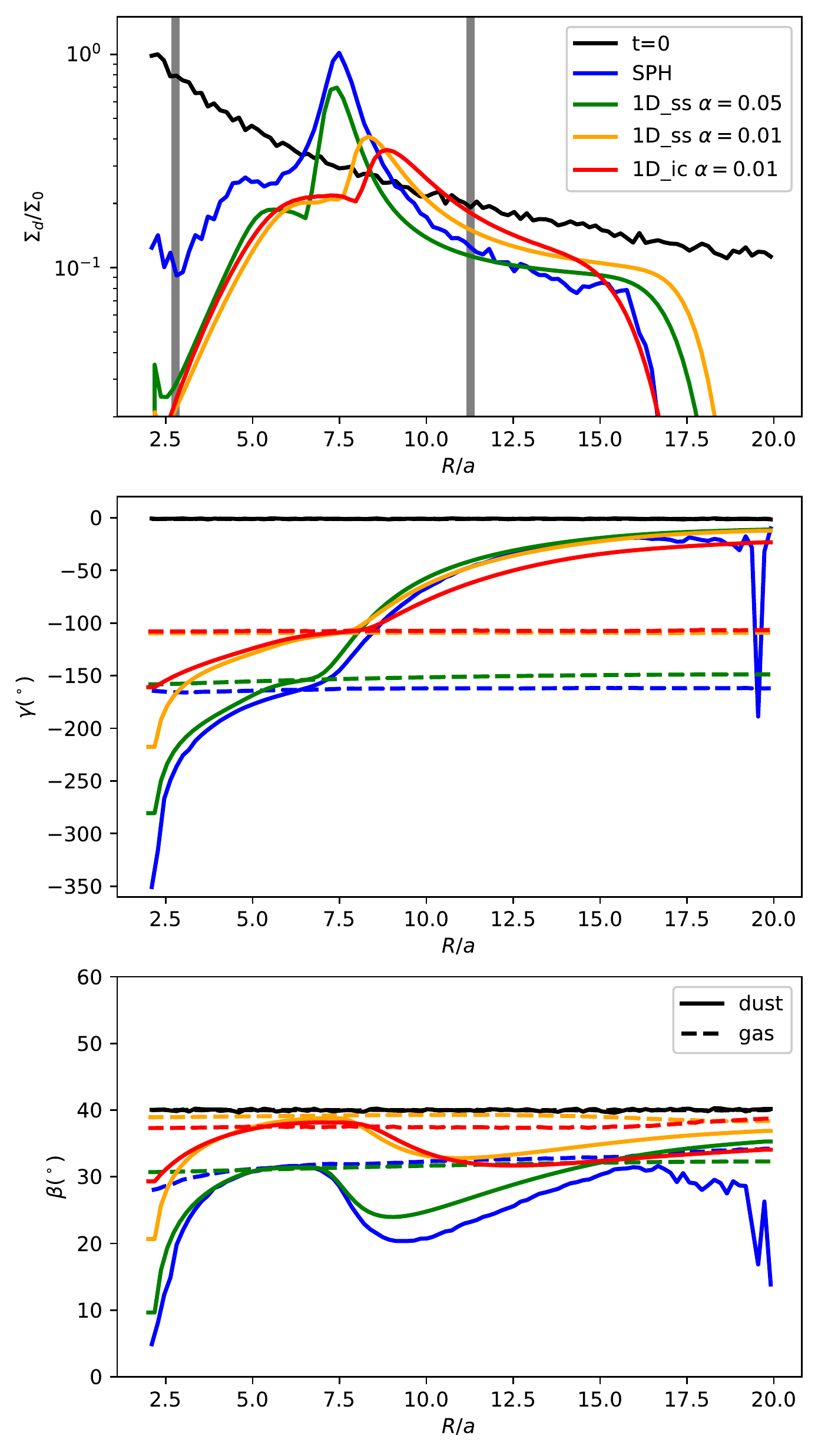}}}

\caption{\label{fig:SPHvRing}
Comparison between the fiducial SPH simulation with corresponding runs using the 1D code at $t=2000$ binary orbits. The vertical grey lines show the locations of the dust rings as predicted by the projection model \citep{LongariniEtal2021}. Panels show radial profiles for the dust surface density (top), twist angle (middle), and tilt angle (bottom).}
\end{figure} 

In Figure~\ref{fig:SPHvRing} we directly compare the SPH results of our fiducial run with our 1D code after 2000 binary orbits. The 1D code qualitatively reproduces all the features in the surface density, twist angle, and tilt angle radial profiles. The twist and tilt angles profiles for both gas and dust show that the $\alpha_\mathrm{d}=0.05$ provides a much closer agreement with SPH than the $\alpha_\mathrm{d}=0.01$ cases, hinting that our simulations indeed suffer from excess dissipation.  We also see that mimicking the steady state SPH gas surface density profile in the 1D code (denoted by $\mathrm{1D}_{\mathrm{ss}}$) enhances the agreement over using a pure power law (denoted by $\mathrm{1D}_{\mathrm{ic}}$). Quantitatively, the RMS error in the dust surface density profiles obtained by the 1D runs compared to that of the SPH simulation are 0.09, 0.17, and 0.175 for the green, orange, and red curves; respectively. This shows that our new dust evolution model faithfully reproduces the dust density bumps caused by the traffic jams, especially when the gas model provides  accurate tilt and twist angles profiles. We note that a lower $\alpha_\mathrm{d}$ only delays the evolution of the density bumps, due to the slower twist angle evolution, but does not inhibit them from forming. While we do not seek to establish the source(s) of the observed excess dissipation in this paper, we propose using this setup (even with gas only) as a method to measure disc visosity in SPH simulations, in addition to the usual `spreading ring' setup (for example \citealt{LodatoPrice2010}, and \citealt{DunhillEtal2013}).

The vertical grey lines in the top panel of Figure~\ref{fig:SPHvRing} show the locations of the inner and outer dust rings predicted by the projection model\footnote{assuming a surface density profile with an exponential cutoff. If instead we assume a pure power law, the projection model gives an outer dust ring radius that is significantly too large ($R/a\simeq14.5$)} presented in \citealt{LongariniEtal2021}. The projection model over(under)-estimates the location of the outer(inner) dust rings, compared to the results from SPH and our new 1D model. This is to be expected since the projection model does not take into account the effects of the drag torque due to the component of the gas velocity normal to that of the dust, which has important consequences on the evolution of the dust disc angular momentum, and especially its tilt profile evolution. In Section~\ref{sec:discuss}, we will discuss the evolution of the tilt profile and how it explains some of the discrepancy between the projection model on one hand, and SPH and 1D model on the other.

\subsection{Different Dust Grain Sizes}

\begin{figure}
\resizebox{85.0mm}{!}{\mbox{\includegraphics[angle=0]{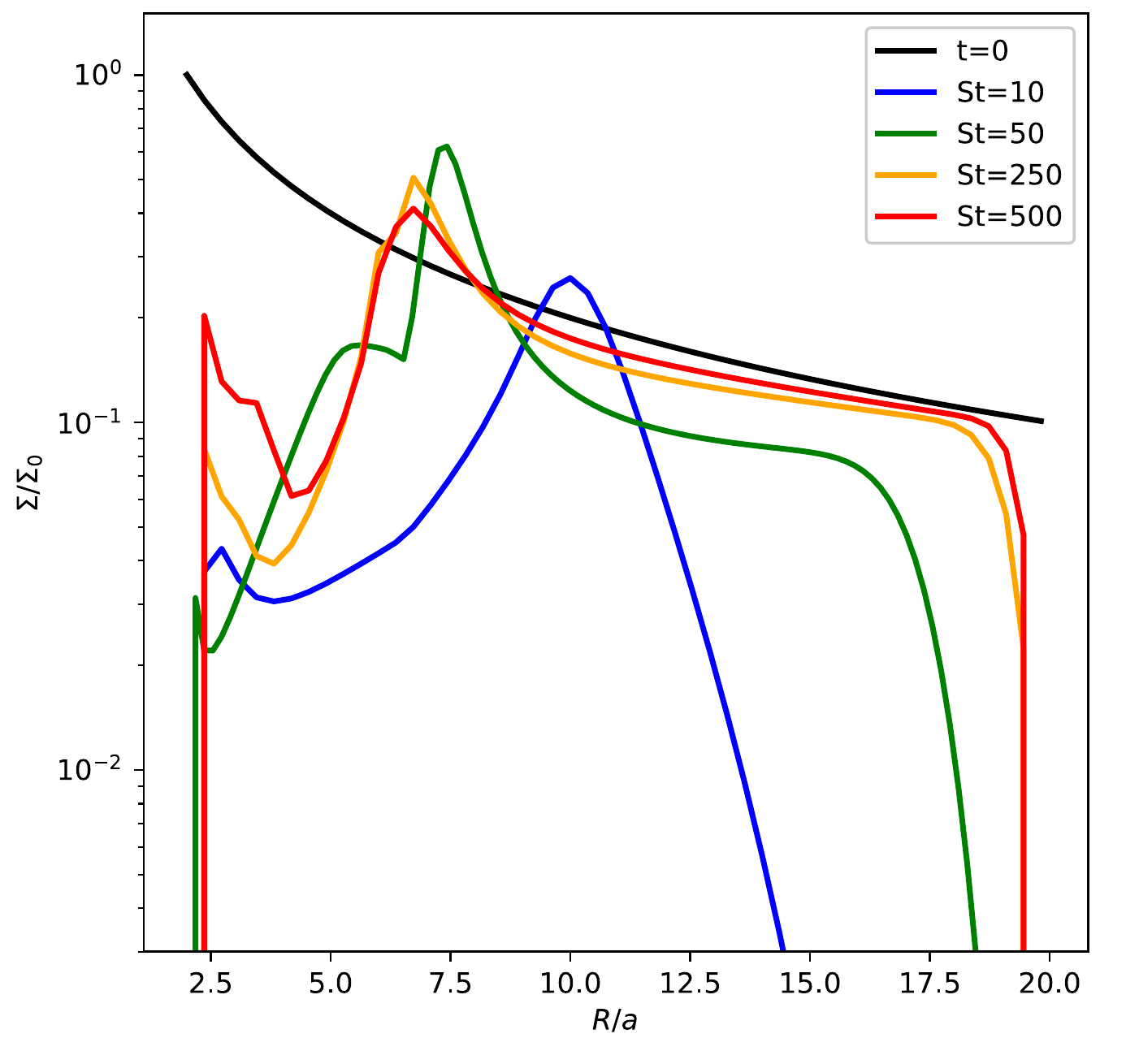}}}

\caption{\label{fig:different_st}
Dust surface density profiles obtained using our 1D code for 4 different dust sizes, for which St=$10, 50, 250, 500$. }
\end{figure}

In section~\ref{sec:SPH} we presented a suite of SPH simulations where we fixed the dust (average) St~$\sim50$ and used our computational resources to vary disc and binary parameters. Here we use our 1D model to investigate the effects of varying St. Figure~\ref{fig:different_st} shows the surface density profiles for 4 different St (10, 50, 250, 500) corresponding to dust sizes of (40{\textmu}m, 200{\textmu}m, 1mm, 2mm) at $t=2000$ binary orbits and fiducial disc and binary parameters. We notice that the drag torque truncates the disc at a radius that increases with St number, since higher St means stronger resistance to the drag torque. We also see that the outer traffic jam radius moves inwards with increasing St. This is explained by the fact that the slope of the dust twist angle profile is expected to be shallower for lower St, and hence the difference in twist angles between gas and dust that is sufficient to create a traffic jam will occur at a larger radius (further out from the co-precession radius).

\section{Discussion}
\label{sec:discuss}
Our new dust angular momentum evolution equation, coupled with gas in a simple 1D implementation, is able to reproduce the location and, to a lesser extent, the amplitude of the traffic jams seen in the SPH simulations. It also reproduced the enhanced radial drift and resultant outer truncation radius, as well as the twist and tilt angles radial profiles. This confirms the theory of traffic jam formation: the drag torque resulting from the different precession frequencies between dust and gas redistributes the dust angular momentum. This redistribution occurs in such a way that the headwind felt by the dust, and hence the radial drift velocity, increases with distance from the co-precession radius. The radial drift is thus minimised at 2 locations bracketing the co-precession radius, where the traffic jams form. The inner traffic jam forms first, since precession is faster, but eventually the outer traffic jam surpasses it as it is closer to the mass reservoir drifting inwards from regions with high radial drift velocity.

The outer traffic jam is long-lived and provides a significant enhancement in dust to gas ratio, up to an order of magnitude increase from the initial value in some of our simulations, which may have significant implications for planetesimal formation, as well as the formation of sub-structures in protoplanetary discs in inclined circumbinary discs. The increased radial drift at large radii, and subsequent disc outer truncation, can have important consequences for disc evolution and may leave a distinctive signal in circumbinary discs population studies.

An important ingredient in the formation of these dust traffic jams is the increased headwind velocity seen by the dust due to the difference in tilt and twist angles between gas and dust. This increases the dust radial drift velocity compared to the co-planar case (see Section~\ref{Sec:DTorTJ}), providing a source of inward drifting material for the traffic jams where the radial velocity is minimised. This is why Figure~\ref{fig:beta} shows the amplitude of the dust pile-ups, as well as the radial drift in the outer disc, increasing with initial inclination angle. Our dust evolution equations predict this effect accurately, as shown in Figure~\ref{fig:SPHvRing}. In this way, our model provides an improvement over the projection model in \citealt{LongariniEtal2021} since it not only predicts the location of the traffic jams, but also the amplitude and indeed whether such traffic jams will develop at all, as well as the evolution of the tilt and twist angles profiles.

\subsection{Morphology}
Figures~\ref{fig:ficucial_dens_dust},~\ref{fig:ecc0.3_dens}, and~\ref{fig:ecc0.6_dens} show that the outer traffic jam always start as two spirals/arcs structures before evolving into a ring at later times. In fact, even at the stage where a circular ring has formed, we can still see that the brightness is not axisymmetric, and the two arcs are still discernible even at later times within the ring (e.g., Figure~\ref{fig:ficucial_dens_dust}). A close examination of Figures~\ref{fig:ficucial_dens_dust} and~\ref{fig:ficucial_dens_gas} shows that the arcs form around the region where the dust and gas rings intersect. As the outer traffic jam starts forming, it is first manifested where the dust crosses the gas mid-plane, as dust particles undergo further compression. This is akin to dust settling. We therefore predict that the spirals/arcs structures will be shorter lived or become less prominent as St approaches unity. We will investigate this prediction in an upcoming paper.

In Figure~\ref{fig:SPHvRing} we notice that the 1D model tends to underestimate the amplitude of the traffic jams. This is likely due to a combination of the vertical averaging of the gas and dust density, as well as the model's inability to capture non-axisymmetric over-densities. 

\subsection{Tilt Evolution}
Normally, we would expect the dust tilt for uncoupled grains to be constant, or, for weakly coupled grains, a monotonically increasing radial profile as the inner disc dissipates energy more efficiently through drag. The tilt profiles shown in this paper exhibit a different behaviour: while the tilt generally increases with radius as expected, in the vicinity of the co-precession radius there seems to be an effect other than dissipation that increases the tilt sharply until it reaches the tilt angle of the gas just inside of the co-precession radius, and decreases it just outside of that radius. Here we explain this phenomenon as a result of the gas drag torque on the dust.

The co-precession radius divides the disc into two regions: an outer region were the gas precession is leading the dust, and an inner region where the gas precession is lagging behind that of the dust. This, combined with the fact that precession around the binary angular momentum vector is always retrograde \citep{AlyEtal2015}, dictates that the gas will torque the dust towards the binary plane (decreasing the dust tilt) outside of the co-precession radius, and away from the binary plane (increasing the dust tilt) inside of the co-precession radius. The effect of this torque component on the dust tilt is most effective close to the co-precession radius where the gas and dust orbits intersect the most and the coupling is enhanced. This explains the peculiar tilt profile seen in Figures~\ref{fig:fiducial_profile},~\ref{fig:ecc_profile}, and~\ref{fig:SPHvRing}.

The projection model developed in \citealt{LongariniEtal2021} considers the geometrical reduction in the dust orbital speed when projected on the gas plane, and predicts dust pile-ups at the two radii when such reduction matches the gas sub-Keplerianity due to pressure support. In order to do this they assume that the gas shares the dust tilt angle throughout the disc. In this paper we showed that this assumption is valid only inside of (and close to) the co-precession radius. The decrease in dust tilt angle just outside the co-precession radius effectively means that the projection model will tend to overestimate the radius of the outer traffic jam, as we can see in Figure~\ref{fig:SPHvRing}, since this discrepancy between the gas and dust tilt is not accounted for in their model. Another consequence of this decrease in dust tilt angle just outside of the co-precession radius is that the radius of the minimum radial drift (and hence the location of the outer traffic jam) moves inwards (closer to the co-precession radius). This can be seen in Figures~\ref{fig:fiducial_profile} and~\ref{fig:beta}.

\subsection{Dust Trap or Traffic Jam?}
\label{Sec:DTorTJ}
Dust traps are regions of pressure maxima where the gas orbits at Keplerian speeds due to the lack of pressure gradient, and the dust gets trapped because it experiences no headwind \citep{NakagawaEtal1986}. There are many mechanisms that cause these dust traps, and they can be radial or azimuthal. In all cases, a dust trap must be able to locally increase the dust to gas ratio. \citet{LongariniEtal2021} referred to the phenomenon studied here as a `Dynamical Dust Trap', with the qualifier `Dynamical' to emphasise the lack of a gas pressure/density bump and that the effect is purely due to the dust speed matching that of the gas because of projection effects. In their projection model, the headwind, and thus the radial drift velocity, completely vanish at two locations around the co-precession radius, causing dust to be locally trapped.

\begin{figure}
\resizebox{85.0mm}{!}{\mbox{\includegraphics[angle=0]{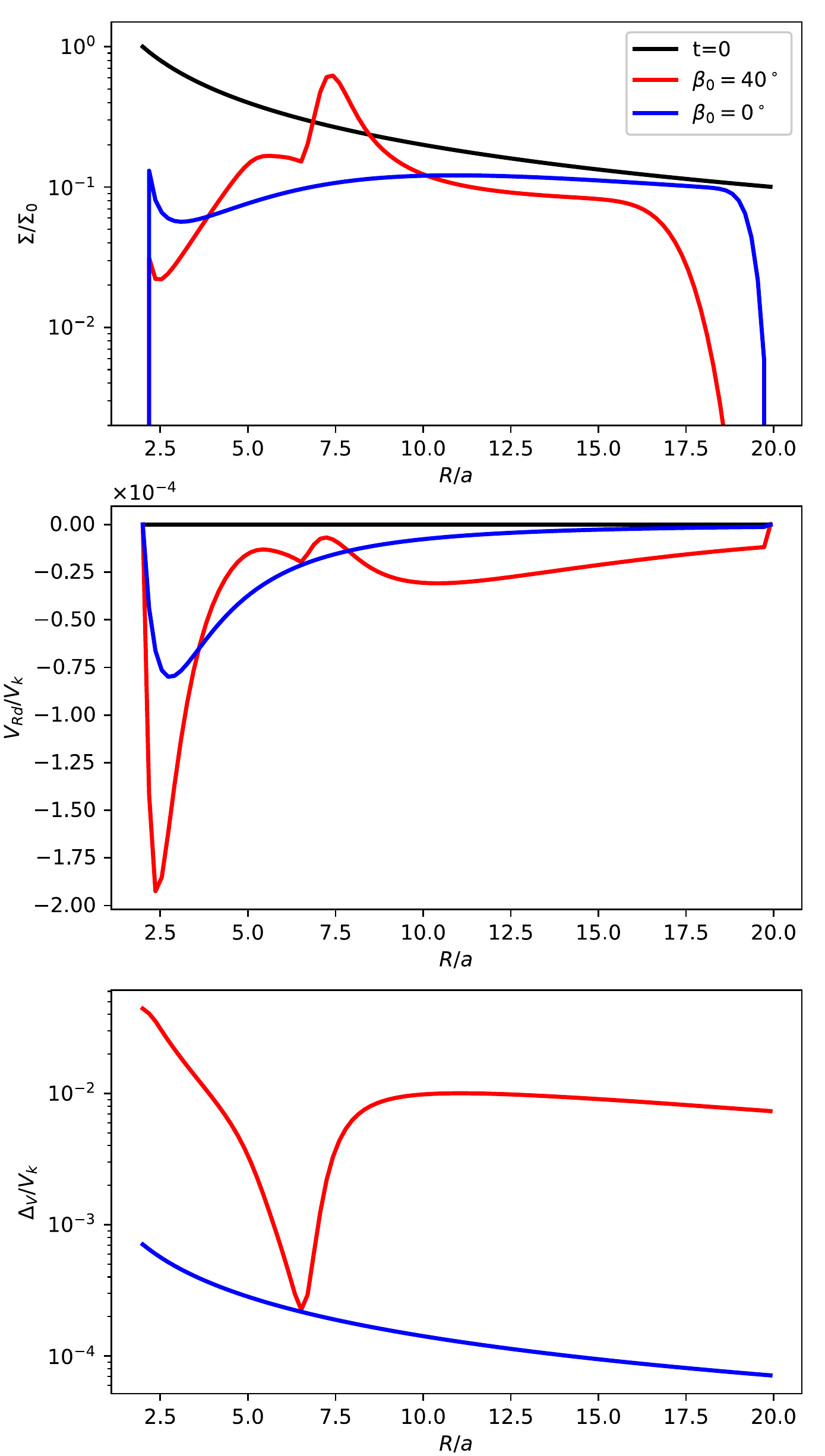}}}

\caption{\label{fig:vr}
Radial profiles comparison for dust surface density (top), radial drift (middle), and headwind velocity (bottom) between the fiducial parameter set ($\beta_0=40^\circ$ and a co-planar ($\beta_0=0^\circ$) case, produced by the 1D code at t=$2000$ binary orbits).}
\end{figure}

Here we aim to delve deeper into the behaviour of the headwind and radial drift velocity in order to assess the most accurate terminology for this phenomenon. In Figure~\ref{fig:vr}, we compare the dust surface density (top), radial drift velocity (middle), and headwind speed (bottom) radial profiles of our fiducial simulation (initial inclination of $40^\circ$) and a co-planar disc around a binary with otherwise identical parameters, simulated with our 1D model and shown here after $2000$ binary orbits. The radial drift and headwind speeds are normalised by the Keplerian velocity. We notice that for the inclined disc, the radial drift velocity is minimised at two locations where the dust piles-up. However, the radial drift never completely vanishes, but rather only gets locally minimised. In fact, the magnitude of the two local minimum radial drift velocities for the inclined case is comparable to the radial drift in the outer disc of the co-planar case. What is more significant than the value of the minimum radial drift velocity is the steep gradient in its profile before the locations where it is minimised. The dust is not trapped, but rather slows down on its inward radial drift, causing a clear traffic jam.

We also see that the headwind speed felt by the dust increases by two orders of magnitude for the inclined disc case, which causes the increase in radial drift in most of the disc and drives the traffic jams. However, at the co-precession radius it decreases rapidly as expected, matching the value for the the co-planar case. 

We therefore note that the inclusion of the drag torque on the dust shows that neither the radial drift velocity, nor the headwind speed, completely vanish as predicted by the projection model. The dust radial drift inwards generally increases everywhere, but is slowed down around (but not exactly at) the co-precession radius, creating dust traffic jams. The dust to gas ratio is enhanced locally at the outer traffic jam. We caution that the term `traffic jams' has been used in the literature to refer to azimuthal clustering of both dust and gas in non co-orbital structures (e.g., eccentric discs, \citealt{AtaieeEtal2013}). These do not cause an enhancement in dust to gas ratio as they form when both gas and dust slow down at the apocentre. As shown throughout this paper, our `dust traffic jams', in contrast, cause significant local increase in dust to gas ratio.
\section{Conclusion}
\label{sec:conclusion}
We investigated the theory of formation of dust traffic jams in inclined circumbinary discs. First, we presented results from SPH simulations showing that these dust traffic jams robustly form for marginally coupled dust particles due to the difference in precession profiles between gas and dust. We showed that dust form long lived sub-structures of enhanced density and dust to gas ratio. These structures initially take the shape of arcs/spirals and later resemble a ring, with hints of azimuthal density dependence persisting until the end of the simulations. We performed a parameter sweep and showed that the traffic jams become more prominent for higher disc initial inclination and/or larger binary eccentricities precessing in the polar regime, with an increase in dust-to-gas ratio up to an order of magnitude from the initial value. 

We then derived the angular momentum evolution equation for a dust ring that is in general tilted and twisted relative to a gas ring, including the gas drag torque on the dust. We implemented this new model in our 1D code \textsc{RiCo}, coupled with the linearised bending-wave warp propagation equations for the gas, appropriate for protoplanetary discs parameters. We showed that our 1D model obtains an excellent agreement with SPH results, albeit with an adjusted dissipation parameter. We investigated the effect of increasing St on the dust surface density profile, showing that while the traffic jam radius moves inward with increasing St, the outer truncation radius moves outward. 

The 1D model allowed us to provide clear radial dust diagnostics and we were able to prove that the cause of the traffic jams at two locations surrounding the co-precession radius is the sharp decrease in both radial drift and headwind velocities, which otherwise increase elsewhere in the disc compared to a reference co-planar case. We speculate that these dust traffic jams may have important implications for planetesimal formation around binaries and may play a role in interpreting observations of disc sub-structures. We will investigate the latter in detail in a follow-up paper.

\section*{Acknowledgments}

We thank the anonymous referee for a thorough report and insightful suggestions which improved the paper.This project has received funding from the European Union’s Horizon 2020 research and innovation programme under the Marie Skłodowska-Curie grant agreement No 823823 (DUSTBUSTERS). HA and JFG acknowledge funding from ANR (Agence Nationale de la Recherche) of France under contract number ANR-16- CE31-0013 (Planet-Forming-Disks) and thank the LABEX Lyon Institute of Origins (ANR-10-LABX-0066) of the Université de Lyon for its financial support within the programme ‘Investissements d’Avenir’ (ANR-11-IDEX-0007) of the French government operated by the ANR. RN acknowledges support from UKRI/EPSRC through
a Stephen Hawking Fellowship (EP/T017287/1). We used OzStar, funded by Swinburne University of Technology and the Australian government. We used \textsc{phantom} \citep{PriceEtal2018} for the hydrodynamic simulations and \textsc{splash} \citep{splash} for rendered images of our simulations

\section*{Data Availability}
The data presented in this article will be shared on reasonable request to the corresponding author. The 1D code \textsc{RiCo} is available upon request to Rebecca Nealon.
\bibliographystyle{mnras} \bibliography{refs}


\appendix
\section{Warp profile in dust and gas}
\label{sec:surface_render}
\begin{figure*}
  \begin{center}
    \resizebox{180.0mm}{!}{\mbox{\includegraphics[angle=0]{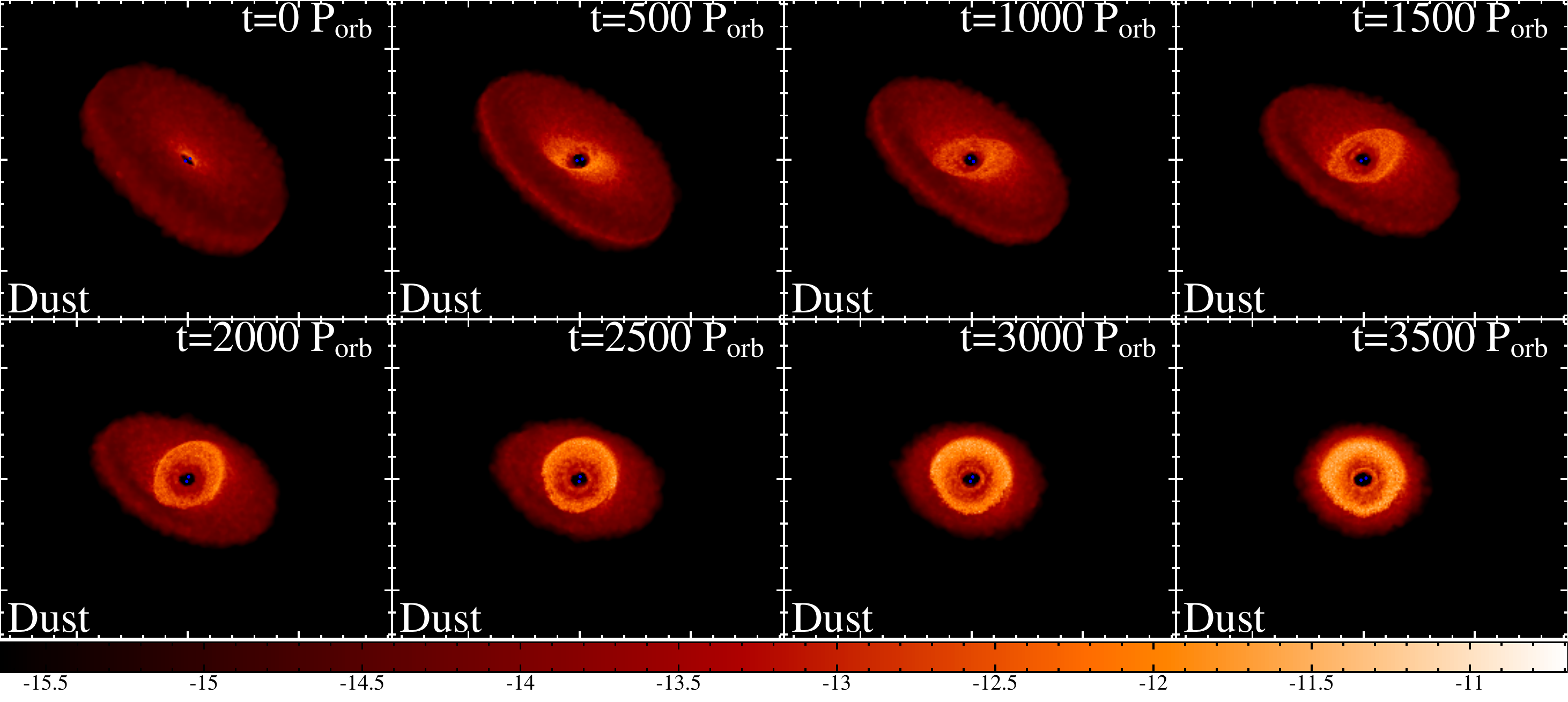}}}
 \caption{\label{fig:dust_surface_render}
 Dust surface rendering at different simulation times for the fiducial simulation ($\beta_0=40^\circ$ and $e_\mathrm{b}=0$).}
 \end{center}
\end{figure*}

\begin{figure*}
  \begin{center}
    \resizebox{180.0mm}{!}{\mbox{\includegraphics[angle=0]{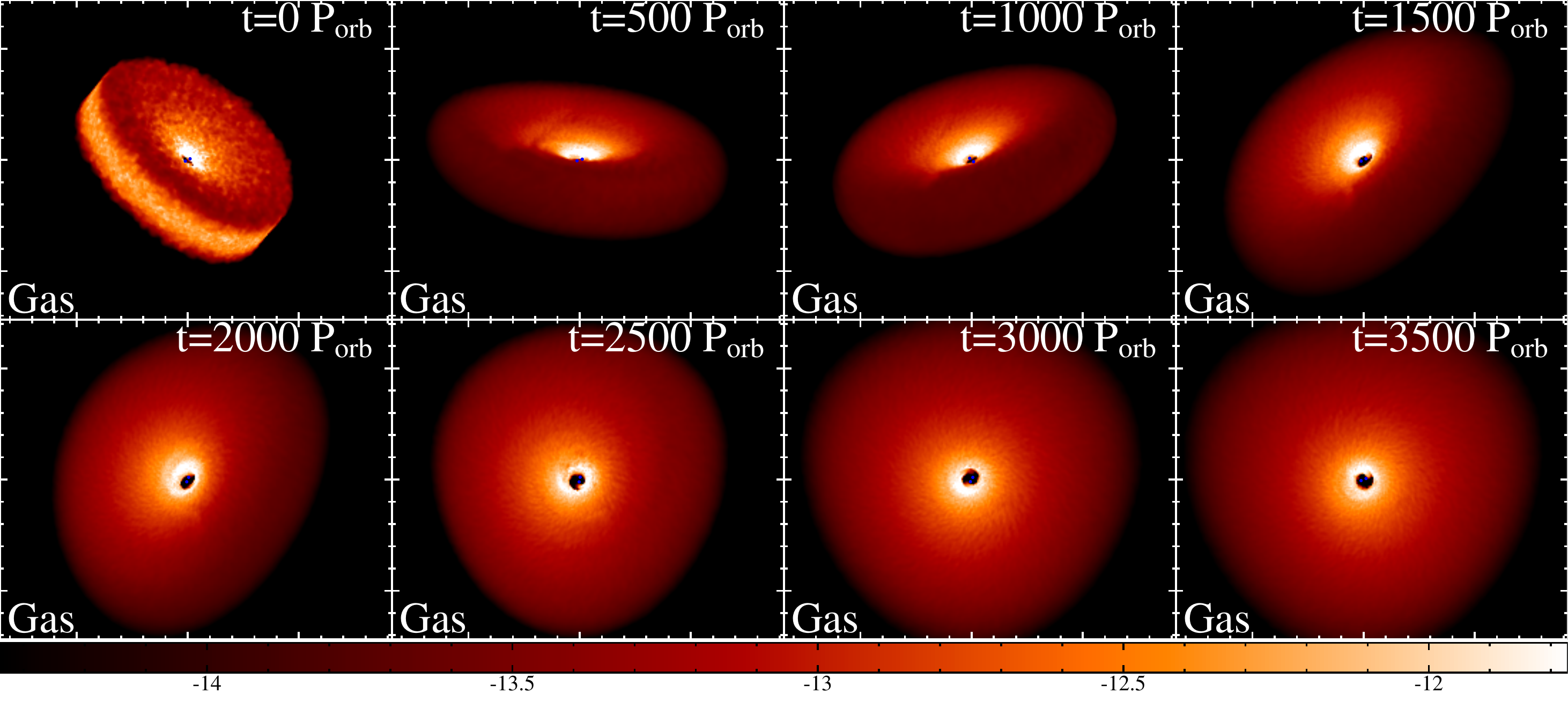}}}
 \caption{\label{fig:gas_surface_render}
 Same as Figure~\ref{fig:dust_surface_render} but for gas.}
 \end{center}
\end{figure*}

The column density maps presented in Figures~\ref{fig:ficucial_dens_dust} and~\ref{fig:ficucial_dens_gas} provide a good view of the morphology of the traffic jams. However, it is somehow difficult to see the structure of the warp. Here, we present in Figures~\ref{fig:dust_surface_render} and~\ref{fig:gas_surface_render} a surface rendering for the dust and gas profiles of the fiducial SPH simulation; respectively. While the surface rendering hides the morphological details of the traffic jams, it provides a better display of the warp in the disc. We can clearly see here that the dust twists and warps throughout the simulation, while the gas precesses almost rigidly and is mostly flat.

\label{lastpage}
\end{document}